\documentclass[submission, Phys]{SciPost}
\usepackage{subcaption}
\usepackage[utf8]{inputenc} % input encodings (allow UTF-8 input)
\usepackage[T1]{fontenc} 	% selecting font encodings (use 8-bit T1 fonts)
\usepackage[english]{babel} % multilingual support (English language/hyphenation)

\usepackage[bitstream-charter]{mathdesign}
\usepackage{geometry} 		% flexible and complete interface to document dimensions
\usepackage{amsmath} 		% math package (American Mathematical Society)
\usepackage{mathtools} 		% math package (fixes various deficiencies of amsmath)
\usepackage{float} 			% floating objects such as figures and tables
\usepackage{graphicx} 		% enhanced support for graphics
\usepackage{tabularx} 		% tabulars with adjustable-width columns
\usepackage{booktabs} 		% professional-quality tables
\usepackage{color, xcolor} 	% foreground and background colour management
\usepackage{pdfpages} 		% inclusion of external multi-page PDF documents
\usepackage{extarrows} 		% extra arrows beyond those provided in amsmath
\usepackage{multirow} 		% create tabular cells spanning multiple rows
\usepackage{multicol} 		% intermix single and multiple columns
\usepackage{enumitem} 		% control layout of itemize, enumerate, description
\usepackage{xspace} 		% define commands that appear not to eat spaces
\usepackage{stackrel} 		% enhancement to the \stackrel command
\usepackage{tikz} 			% create PostScript and PDF graphics
\usepackage{braket} 		% Dirac bra-ket notation
\usepackage{bm} 			% access bold symbols in maths mode
\usepackage{tensor} 		% typeset tensors (tensor-style super- and subscripts)
\usepackage{slashed} 		% slash through characters (Feynman slash notation)
\usepackage{siunitx} 		% SI units package (typesetting values with units)
\usepackage{lastpage} 		% reference last page
\usepackage{cite} 			% improved citation handling
\usepackage[normalem]{ulem} % package for underlining
\usepackage{fontawesome} 	% access to web-related icons
\usepackage{tocloft} 		% control over the typography of the TOC, LOF, and LOT
\usepackage{titlesec} 		% interface to sectioning commands (various title styles)
\usepackage{doi} 			% create correct hyperlinks for DOI numbers
\usepackage{hyperref} 		% hypertext links (handel cross-referencing commands)
\usepackage[most]{tcolorbox} 					% coloured and framed text boxes
\usepackage[nameinlink, capitalize]{cleveref} 	% intelligent cross-referencing
\usepackage[nottoc, notlot, notlof]{tocbibind} 	% add references/index/contents to TOC
\usepackage[ruled, vlined]{algorithm2e} 		% floating algorithm environment
\usepackage{makecell}
\usepackage[makeroom]{cancel}
\usepackage{feynmf}

\makeatletter
\def\BState{\State\hskip-\ALG@thistlm}
\makeatother

\makeatletter
\@ifundefined{pdfoutput}{}{\DeclareGraphicsRule{*}{mps}{*}{}}
\makeatother

\makeatletter
\DeclareRobustCommand*{\bfseries}{%
   \not@math@alphabet\bfseries\mathbf
   \fontseries\bfdefault\selectfont
   \boldmath
}
\makeatother

\hypersetup{
	pdftitle={Advancing Tools for Simulation-Based Inference},
	pdfauthor={Bahl H., Bres\'{o} V., De Crescenzo G., Plehn T.},
	colorlinks=true, 			% false: boxed links, true: colored links
	linkcolor={red!50!black}, 	% color of internal links (sections, pages, etc.)
	citecolor={blue!50!black}, 	% color of citation links (links to bibliography)
	urlcolor={blue!80!black} 	% color of URL links (external links)
} 
\DeclareSymbolFont{usualmathcal}{OMS}{cmsy}{m}{n}
\DeclareSymbolFontAlphabet{\mathcal}{usualmathcal}

\SetArgSty{textnormal}
\SetKwComment{Comment}{{\small\#}~}{}
\SetCommentSty{mycommfont}

\setitemize{itemsep=0pt, parsep=0pt} 				% adjust itemize environment
\setenumerate{itemsep=0pt, parsep=0pt} 				% adjust enumerate environment
\setitemize{itemsep=2pt,topsep=2pt,parsep=0pt,partopsep=0pt,leftmargin=*}
\setenumerate{itemsep=0pt,topsep=2pt,parsep=0pt,partopsep=0pt,labelindent=3pt,leftmargin=*}
\usepackage{amsmath}
 
\usepackage{amsthm} 		% math package (typesetting theorems using AMS style)
\theoremstyle{definition}

\usetikzlibrary{arrows, shapes.geometric, arrows.meta, shapes, decorations.pathreplacing, fit, patterns, patterns.meta,positioning,calc}
\usepackage{tikz} 			% create PostScript and PDF graphics

\definecolor{Rcolor}{HTML}{E99595}
\definecolor{Gcolor}{HTML}{C5E0B4}
\definecolor{Gcolor_light}{HTML}{F1F8ED}
\definecolor{Bcolor}{HTML}{9DC3E6}
\definecolor{Ycolor}{HTML}{FFE699}
\definecolor{Ycolor_light}{HTML}{FFF7DE}
\newcommand{\tikznode}[2]{%
\ifmmode%

\tikz[remember picture,baseline=(#1.base),inner sep=0pt] \node (#1) {$#2$};%
\else
\tikz[remember picture,baseline=(#1.base),inner sep=0pt] \node (#1) {#2};%
\fi}

\tikzstyle{expr} = [rectangle, rounded corners=0.3ex, minimum width=1.5cm, minimum height=1cm, text centered, align=center, inner sep=0, fill=Ycolor, font=\large, draw]

\tikzstyle{small_cinn} = [double arrow, double arrow head extend=0cm, double arrow tip angle=130, inner sep=0, align=center, minimum width=1.1cm, minimum height=0.5cm, fill=Rcolor, draw]

\tikzstyle{small_cinn_black} = [small_cinn, minimum height=1.5cm, fill=black]

\tikzstyle{transformer} = [rectangle, rounded corners, minimum width=6cm, minimum height=2.4cm, font=\large, fill=Gcolor_light, draw]

\tikzstyle{attention} = [rectangle, rounded corners=0.3ex, minimum width=5.5cm, minimum height=1.2cm, align=center, fill=Gcolor, draw, font=\large]

\tikzstyle{transformer_huge} = [rectangle, rounded corners, minimum width=8.5cm, minimum height=2.4cm, font=\large, fill=Gcolor_light, draw]

\tikzstyle{attention_huge} = [rectangle, rounded corners=0.3ex, minimum width=8cm, minimum height=1.2cm, align=center, fill=Gcolor, draw, font=\large]

\tikzstyle{txt_huge} = [align=center, font=\Huge, scale=2]
\tikzstyle{txt} = [align=center, font=\LARGE, minimum height=1cm]

\tikzstyle{arrow} = [thick,-{Latex[scale=1.0]}, line width=0.2mm, color=black]
\tikzstyle{line} = [thick, line width=0.2mm, color=black]
\tikzstyle{bkg} = [minimum width=22cm, minimum height=7.5cm, font=\large, fill=white]

\tikzstyle{encoder_black} = [trapezium, fill=black, rotate=270, minimum height=4cm, minimum width=3.2cm,align=center,draw]
\tikzstyle{encoder} = [trapezium, fill=Bcolor, rotate=270, minimum height=4cm, minimum width=3cm,align=center,draw]
\tikzset{trapezium stretches=true}
\tikzstyle{cinn} = [double arrow, double arrow head extend=0cm, double arrow tip angle=130, shape border rotate=90, inner sep=0, align=center, minimum width=2.1cm, minimum height=2.3cm, fill=Rcolor, draw,font=\LARGE]
\tikzstyle{cinn_black} = [cinn, minimum height=2.5cm, fill=black]
\tikzstyle{crc} = [circle, rounded corners=0.3ex, minimum width=1.5cm, minimum height=1cm, text centered, align=center, inner sep=0, fill=white, font=\LARGE, draw] 
\definecolor{red_cb}{HTML}{e41a1c}
\definecolor{blue_cb}{HTML}{377eb8}
\definecolor{green_cb}{HTML}{4daf4a}
\definecolor{purple_cb}{HTML}{984ea3}
\definecolor{orange_cb}{HTML}{ff7f00}

\definecolor{EmeraldGreen}{HTML}{1ea78d}
\definecolor{EnglishRed}{HTML}{b02427}
\newcommand{\XXLangle}{\biggl\langle}
\newcommand{\XXRangle}{\biggr\rangle}

\newcommand{\qqquad}{\qquad\quad}

\newcommand\one{\leavevmode\hbox{\small1\normalsize\kern-.33em1}}
\newcommand{\order}{\mathcal{O}} 			% order
\newcommand{\normal}{\mathcal{N}}
\newcommand{\lag}{\mathscr{L}}
\newcommand{\loss}{\mathcal{L}}

\newcommand{\mat}{\mathcal{M}}

\newcommand{\madgraph}{\textsc{MadGraph5\_aMC@NLO}\xspace}

\newcommand{\pythia}{\textsc{Pythia8}\xspace}
\newcommand{\fastjet}{\textsc{FastJet}\xspace}

\newcommand{\delphes}{\textsc{Delphes}\xspace}

\newcommand{\madspin}{\textsc{MadSpin}\xspace}
\newcommand{\madminer}{\textsc{MadMiner}\xspace}
\newcommand{\arXiv}[2][]{%
	\ifthenelse{\equal{#1}{}}%
	{\href{http://arxiv.org/abs/#2}{arXiv:#2}}%
	{\href{http://arxiv.org/abs/#2}{arXiv:#2~[#1]}}}

\newcommand{\gev}{\text{GeV}}
\newcommand{\tev}{\text{TeV}}

\newcommand{\sig}{\text{sig}}
\newcommand{\bkg}{\text{bkg}}
\newcommand{\ifb}{\ensuremath{\text{fb}^{-1}} }

\def\slashchar#1{\setbox0=\hbox{$#1$}           % set a box for #1
   \dimen0=\wd0                                 % and get its size
   \setbox1=\hbox{/} \dimen1=\wd1               % get size of /
   \ifdim\dimen0>\dimen1                        % #1 is bigger
      \rlap{\hbox to \dimen0{\hfil/\hfil}}      % so center / in box
      #1                                        % and print #1
   \else                                        % / is bigger
      \rlap{\hbox to \dimen1{\hfil$#1$\hfil}}   % so center #1
      /                                         % and print /
   \fi}

\def\mathswitchr#1{\relax\ifmmode{\mathrm{#1}}\else$\mathrm{#1}$\xspace\fi}
\def\mathswitch#1{\relax\ifmmode#1\else$#1$\xspace\fi}

\newcommand{\ETmiss}{\ensuremath{\slashed{E}_T}\xspace}

\DeclarePairedDelimiterX{\infdivx}[2]{[}{]}{%
  #1\;\delimsize\|\;#2%
}

\graphicspath{{./figs/}}
\begin{document}

\begin{center}{\Large \textbf{
Advancing Tools for Simulation-Based Inference
}}\end{center}

\begin{center}
  Henning Bahl\textsuperscript{1}, 
  Victor Bres\'{o}-Pla\textsuperscript{1},
  Giovanni De Crescenzo\textsuperscript{1}, 
  and Tilman Plehn\textsuperscript{1,2}
\end{center}

\begin{center}
{\bf 1} Institut f\"ur Theoretische Physik, Universit\"at Heidelberg, Germany \\
{\bf 2} Interdisciplinary Center for Scientific Computing (IWR), Universit\"at Heidelberg, Germany
\end{center}

\begin{center}
\today
\end{center}

% For convenience during refereeing: line numbers
%\linenumbers

\section*{Abstract}

We study the benefit of modern simulation-based inference to
constrain particle interactions at the LHC. We explore
ways to incorporate known physics structures into likelihood estimation, 
specifically morphing-aware estimation and derivative
learning. Technically, we introduce a new and more efficient smearing 
algorithm, illustrate how uncertainties can be 
approximated through repulsive ensembles, and show how 
equivariant networks can improve likelihood estimation.
After illustrating these aspects for a toy model, we 
target di-boson production at the LHC and find that 
our improvements significantly increase 
numerical control and stability.

\vspace{10pt}
\noindent\rule{\textwidth}{1pt}
\tableofcontents\thispagestyle{fancy}
\noindent\rule{\textwidth}{1pt}
%\vspace{10pt}

\clearpage
%%%%%%%%%%%%%%%%%%%%%%%%%%%%%%%%%%%%%%%%%%%%%%%%%%%
\section{Introduction}
\label{sec:intro}

Since the Higgs discovery, the experimental and theoretical
communities have turned the LHC into the first precision hadron
collider in history. Measurements and simulations of LHC event kinematic patterns are
high-dimensional. This is an advantage because a lot of information
can be used to test the underlying theory.  However, it challenges
traditional methods, which rely on at most few-dimensional
histograms as summary statistics. Modern machine learning techniques can be used to overcome this bottleneck~\cite{Butter:2022rso,Plehn:2022ftl},
specifically unbinned likelihood (ratio) estimation~\cite{Cranmer:2019eaq}.

To interpret LHC measurements systematically, we need to (i) optimally
extract information from observed and simulated events, and (ii)
publish this information in a useful manner.  In both cases, total or
fiducial rate measurements alone are not enough. Instead, we need to
analyze and publish full likelihoods or likelihood ratios, including
an accurate modeling of systematic uncertainties and their
correlations.  One approach to optimal inference is the so-called
matrix element method, enabled by modern machine
learning~\cite{Butter:2022vkj,Heimel:2023mvw}.  Alternatively,
high-dimensional and unbinned likelihood ratio estimation can be
achieved using simulation-based inference (SBI), which is also based on modern machine learning~\cite{Brehmer:2018eca,Brehmer:2018kdj,Brehmer:2018hga,Kong:2022rnd}. SBI was first implemented
for LHC purposes in the public \madminer{}
tool~\cite{Brehmer:2019xox}. The promise of this analysis approach has
been documented by several phenomenological studies, for instance
targeting $VH$~\cite{Brehmer:2019gmn},
$ZH$~\cite{GomezAmbrosio:2022mpm}, $t\bar
t$~\cite{GomezAmbrosio:2022mpm}, and Higgs pair
production~\cite{Mastandrea:2024irf} using the effective theory
extension of the Standard Model (SMEFT)~\cite{Brivio:2017vri}, or
CP-phases in $t\bar{t}H$~\cite{Barman:2021yfh,Bahl:2021dnc}
and $WH$ production~\cite{Barrue:2023ysk}.

One outcome of these phenomenological analyses is a set of open questions
concerning the numerical stability of likelihood ratio estimation,
especially its scaling towards the fully exclusive phase space and
high-dimensional theory space. Here, we can use our physics
understanding of LHC phase space and of the perturbative structure of
the underlying quantum field theory.

This is for instance evident in SMEFT, a theoretical description which 
allows us to combine rate
and kinematic information from different
processes~\cite{Butter:2016cvz,Brivio:2019ius,Elmer:2023wtr}. In
SMEFT, the scattering matrix element is expressed as a truncated Taylor series in the Wilson
coefficients. This means that we can extract first
and second derivatives rather than the full likelihood
ratio~\cite{Chatterjee:2021nms,Chatterjee:2022oco,GomezAmbrosio:2022mpm},
including systematic
uncertainties~\cite{Schofbeck:2024zjo}. Alternatively, this structure
can also be exploited by learning the likelihood ratio at a set of
basis points~\cite{Brehmer:2018eca,Chen:2023ind}.
    
In this paper, we target some of the known numerical shortcomings of
the standard \madminer tool box. In Sec.~\ref{sec:likelihood_learning}
we discuss different ways of using our physics knowledge about LHC
scattering amplitudes to improve likelihood learning.  The advantages
of morphing-aware sampling and derivative learning are best
illustrated for a toy model in Sec.~\ref{sec:toy_compare}. The
fractional smearing technique introduced in Sec.~\ref{sec:toy_frac}
leads to a critical numerical improvement for our toy model and also
for the $WZ$ application discussed in detail in
Sec.~\ref{sec:lhc}. The benefits from the covariant L-GATr
architecture become relevant once we probe higher-dimensional phase
space, for instance for $WZ$ production at the reconstruction level.

%%%%%%%%%%%%%%%%%%%%%%%%%%%%%%%%%%%%%%%%%%%%%%%%%%%%%%%%%%%%%%%%%%%
\section{Learning the likelihood}
\label{sec:likelihood_learning}

The extraction of likelihood ratios at the LHC involves two phase spaces and their
underlying distributions, $z_p$ for the hard scattering and $x$ for
instance for the (smeared) reconstruction level. Our goal is to
extract the likelihood ratio $r(x|\theta, \theta_0)$, defined as the ratio between the likelihood with free $\theta$ against a fixed reference hypothesis $\theta_0$,
\begin{align}
    r(x |\theta,\theta_0) \equiv \frac{p(x|\theta)}{p(x|\theta_0)} \;.
    \label{eq:def_lr}
\end{align}
There are two main strategies which can be used to achieve this. First of all, we review the structure of the problem. 
\subsubsection*{LHC likelihood factorization}
Following Refs.~\cite{Brehmer:2018eca, Brehmer:2018kdj,Brehmer:2018hga}, we can
assume that our parameters of interest only affect the hard scattering
observables $z_p$, while the parton shower (transferring the $z_p$ to
the shower variables $z_s$), the detector simulation (transferring the
$z_s$ to the detector variables $z_d$), and the forming of the
reco-level observables $x$ out of the $z_d$ are independent
of $\theta$. In this case, we assume that the likelihood factorizes as
\begin{align}
  p(x|\theta) 
  = \int dz_d dz_s dz_p p(x|z_d) p(z_d|z_s) p(z_s|z_p) p(z_p|\theta) 
  =\int dz_p p(x|z_p) p(z_p|\theta)\;, \label{eq:pxtheta_int}
\end{align}
where in the last step $p(z_p|\theta)$ can be evaluated using parton-level event generators:
\begin{align}
  p(z_p|\theta) = \frac{1}{\sigma(\theta)}\frac{d\sigma(z_p|\theta)}{dz_p} \; .
\end{align}
The differential cross-section reads
\begin{align}
  d\sigma(z_p|\theta) 
  = (2\pi)^4 \int dx_1dx_2 d\Phi \frac{f_1(x_1,Q^2)f_2(x_2,Q^2)}{2x_1 x_2 s}|\mat(z_p|\theta)|^2 \;,
\end{align}
where $f_i(x,Q^2)$ are the parton densities depending on the momentum
transfer $Q$ and partonic momentum fractions $x_i$, $\mat(z_p|\theta)$
is the matrix element of the hard scattering and $d\Phi$ is the
phase-space element.
This factorization is important because it allows us to compute numerically the \textit{joint likelihood ratio}, which is simply the ratio of the joint probabilities. The full likelihood ratio is clearly intractable, but the joint likelihood ratio simply reduces to the parton level likelihood ratio:
\begin{align}
    r(x,z_d,z_s,z_p|\theta,\theta_0) := \frac{p(x|z_d)p(z_d|z_s)p(z_s|z_p)p(z_p|\theta)}{p(x|z_d)p(z_d|z_s)p(z_s|z_p)p(z_p|\theta_0)}=\frac{p(z_p|\theta)}{p(z_p|\theta_0)}=:r(z_p|\theta,\theta_0)\;.
    \label{eq:joint_ratio_definition}
\end{align}
This quantity is available from simulation and it can be exploited to make training easier, given we are feeding the NN with more information. We can now present the two main approaches to approximate the quantity of interest $r(x|\theta,\theta_0)$.

%%%%%%%%%%%
\subsubsection*{Likelihood learning via classifier}

By training a discriminator $D(x)$ to separate the effects of the
parameter choices $\theta$ and the reference hypothesis $\theta_0$, we
can extract the likelihood ratio at the reconstruction level. If the
classifier is optimally trained, it has the form
\begin{align}
    D_\text{opt}(x|\theta,\theta_0) = \frac{p(x|\theta_0)}{p(x|\theta_0) + p(x|\theta)}\;.
\end{align}
Then, the likelihood ratio is given by
\begin{align}
    r(x |\theta,\theta_0) = \frac{1 - D_\text{opt}(x|\theta,\theta_0)}{D_\text{opt}(x|\theta,\theta_0)}\;.
\end{align}
This is the CARL method implemented in
\madminer~\cite{Brehmer:2018kdj,Brehmer:2019xox}. 
To learn the classifier we define a learnable
$f_\varphi(x|\theta)$ and study the functional
\begin{align}
    F[f_\varphi]=\int d\theta q(\theta)\left[\int dxp(x|\theta_0)\log(f_\varphi(x|\theta))+\int dxp(x|\theta)\log(1-f_\varphi(x|\theta))\right]\; ,
    \label{eq:BCE_CARL_functional}
\end{align}
where we ignore the second argument $\theta_0$. Minimizing $F$ with respect to $f_\varphi(x|\theta)$ leads to the condition
\begin{align}
    0 = \frac{\delta F}{\delta f_\varphi} 
    \propto \frac{p(x|\theta_0)}{f_\varphi(x|\theta)}-\frac{p(x|\theta)}{1-f_\varphi(x|\theta)} 
\qquad \Leftrightarrow \qquad 
f_\varphi(x|\theta)=\frac{p(x|\theta)}{p(x|\theta_0)+p(x|\theta)}\;.
\end{align}
This method does not make use of parton-level information. To include it, the functional form can be adapted leading to the same minimum for $f_\varphi(x|\theta)$
\begin{align}
    F[f_\varphi]=-\int d\theta q(\theta)\Bigg[
    & \int dx\int dz_p \; p(x|z_p)p(z_p|\theta_0)
    \; \frac{1}{1+r(z_p|\theta,\theta_0)} \; \log f_\varphi(x|\theta) \notag \\
    + & \int dx\int dz_pp(x|z_p)p(z_p|\theta) \; \left(1-\frac{1}{1+r(z_p|\theta,\theta_0)}\right) \; \log(1-f_\varphi(x|\theta))\Bigg]\;.
    \label{eq:BCE_ALICE_functional}
\end{align}
This is the ALICE method implemented in
\madminer~\cite{Stoye:2018ovl}. In principle, it should lead to faster training given the additional parton-level information the network is fed. The ALICE method is implemented through the loss
\begin{align}
    \loss = -\XXLangle \Bigg[& \frac{1}{1+r(z_p|\theta,\theta_0)}\log r_\varphi(x|\theta,\theta_0) \notag \\
    + &\left(1-\frac{1}{1+r(z_p|\theta,\theta_0)}\right)\log(1-r_\varphi(x|\theta,\theta_0)\Bigg] \XXRangle_{x,z_p \sim \frac{p(x|z_p)p(z_p|\theta)+p(x|z_p)p(z_p|\theta_0)}{2};\theta\sim q(\theta)}  \; .
    \label{eq:BCE_ALICE_loss}
\end{align}

%%%%%%%%%%%
\subsubsection*{Likelihood regression}

As an alternative, we can also regress to the reco-level
likelihood using parton-level or hard-scattering information ~\cite{Brehmer:2018eca}. 
While, in general, we cannot evaluate the integral in
Eq.\eqref{eq:pxtheta_int}, we can learn it by using
$p(z_p|\theta)$. For our learnable
$f_\varphi(x|\theta)$, we investigate the functional
\begin{align}
  F[f_\varphi] = \int dx\int d\theta\int dz_p q(\theta)p(x|z_p)p(z_p|\theta)\left[f(z_p|\theta) - f_\varphi(x|\theta)\right]^2\;, \label{eq:Ffunctional}
\end{align}
where $f(z_p)$ is some function of the parton-level
variables. Minimizing $F$ with respect to $f_\varphi(x|\theta)$ leads
to the condition,
\begin{align}
    0 &= \frac{\delta F}{\delta f_\varphi(x|\theta)} 
    \propto \int  \; dz_p p(x|z_p) p(z_p|\theta)\left[f(z_p|\theta) - f_\varphi(x|\theta)\right] \;.
\end{align}
Here, the functional derivative with respect to $f_\varphi(x|\theta)$
effectively removes the integrals over $\theta$ and $x$. Solving for
$f_\varphi(x|\theta)$, we obtain
\begin{align}
    f_\varphi(x|\theta) &= \frac{\int d z_p p(x|z_p)p(z_p|\theta) f(z_p|\theta)}{\int d z_p p(x|z_p)p(z_p|\theta)} = \frac{\int d z_p p(x|z_p)p(z_p|\theta) f(z_p|\theta)}{p(x|\theta)} \;.
\label{eq:lr_trick}
\end{align}
We can use this method for the joint likelihood ratio, such that
\begin{align}
 f(z_p|\theta)
    \equiv r(z_p|\theta_0, \theta) = \frac{1}{r(z_p|\theta, \theta_0)} 
  = \frac{p(z_p|\theta_0)}{p(z_p|\theta)}
  \label{eq:f_choice}
\end{align}
and
\begin{align}
  f_\varphi(x|\theta) \equiv \frac{1}{r_\varphi(x|\theta,\theta_0)} \approx \frac{1}{r(x|\theta,\theta_0)} \; .
\end{align}
This way we learn the likelihood ratio at the reco-level. This
minimization can be implemented through the loss
\begin{align}
    \loss &= 
    \XXLangle \left[ r(z_p|\theta,\theta_0)- r_\varphi(x|\theta,\theta_0) \right]^2 \XXRangle_{x,z_p \sim p(x|z_p)p(z_p|\theta);\theta\sim q(\theta)}  \; .
    \label{eq:MSE_ratio_loss}
\end{align}
Each training event includes the parton-level momenta $z_{p,i}$, the
reco-level observables $x_i$, and the theory parameters $\theta_i$. In
addition, we attach the parton-level likelihood ratio as a label. As
we will see in Sec.~\ref{sec:derivative_learning}, one can regress on
a range of possible targets using this MSE, all relying on the
knowledge of the parton-level likelihood ratio.

The exact form of the loss is important. For example, if taking a
different exponent or computing the mean-squared-error w.r.t.\ $\log
r(z_p|\theta,\theta_0)$, $r_\varphi(x)$ will in general not converge
to $r(x|\theta,\theta_0)$. Linear operations, like scaling the targets
to have mean of zero and a standard deviation of one, will not affect
the convergence, since multiplication and addition commute with the
integral. This can be seen by focusing on Eq.~\eqref{eq:lr_trick}: we can see that $f(z_p|\theta)$ needs to appear linearly in the integral in order for the cancellation between numerator and denominator $p(z|\theta)$ to happen (see Eq.~\eqref{eq:f_choice}).

The most straightforward way to learn the $\theta$-dependence of $r$
is to sample the training data from various $\theta$-hypotheses and to
condition the neural network on $\theta$. Since directly sampling from
a large set of theory hypotheses is not feasible in practice, the
samples must be generated using morphing techniques, as discussed
below. Alternatively, the theory parameter dependence can be directly
imprinted by splitting the likelihood ratio into several functions
approximated by separate neural networks.

%%%%%%%%%%%%%%%%%%%%%%%%%%%%%%%%%%%%%%%%%%%%%%%%%%%%%%%%%%%%%%%%%%%
\subsection{Morphing}
\label{sec:morphing}

All modern Monte Carlo generators combine phase-space sampling with
additional dimensions, for instance helicities or color. Naturally, we
can do the same with model parameters to produce an efficient
training dataset to learn likelihood
ratios~\cite{Brehmer:2018eca}. For instance, for BSM physics described by an
effective Lagrangian up to dimension six,
\begin{align}
    \lag_\text{SMEFT} 
    = \lag_\text{SM} + \sum_{i}\frac{c_i}{\Lambda^2} \; O_i 
    \equiv \lag_\text{SM} + \sum_{i}\theta_i \; O_i \;,
\end{align}
the differential cross-section for the hard scattering has the form 
\begin{align}
  |\mat(z_p|\theta)|^2 = |\mat_\text{SM}(z_p)|^2 + \theta_i |\mat_i(z_p)|^2 + \theta_i \theta_j |\mat_{ij}(z_p)|^2 \; .
  \label{eq:mat_factorize}
\end{align}
Even though we use SMEFT as our example, the same form applies
to any polynomial dependence on couplings of interest. This form also holds beyond leading order.

Using Eq.\eqref{eq:mat_factorize}, we can factorize the differential and total partonic
cross-sections as
\begin{align}
  \frac{d \sigma(z_p|\theta)}{dz_p} \equiv w_a(\theta) \; f_a(z_p) 
  \qquad \text{and} \qquad 
  \sigma(\theta) = w_a(\theta) \int dz_p \; f_a(z_p) \; .
  \label{eq:dsigma_factorize}
\end{align}
This separates the $\theta$-dependence in $w_a(\theta)$ and the
$z_p$-dependence in $f_a(z_p)$. For a single $\theta$, the sum over
$a$ includes three terms, the constant SM term, the linear, and the
quadratic BSM contributions.  The normalized probability distribution
is then
\begin{align}
  p(z_p|\theta)
  &= \frac{1}{\sigma(\theta)} \; \frac{d\sigma(z_p|\theta)}{dz_p}
  = \frac{w_a(\theta)}{\sigma(\theta)} \; f_a(z_p) \; .
\label{eq:fact_pz}
\end{align}
For a suitable set of points $\theta_i$ and a set of $a$-terms, this
matrix relation can be inverted
\begin{align} 
  p(z_p|\theta_i)
  = \frac{w_a(\theta_i)}{\sigma(\theta_i)} \; f_a(z_p) 
  \qquad \Leftrightarrow \qquad 
    f_a(z_p) = \left[ \frac{w_a(\theta_i)}{\sigma(\theta_i)}\right]^{-1}p(z_p|\theta_i) \; .
\end{align}
This way, we can use basis densities $p(z_p|\theta_i)$ to reconstruct the
full dependence on $\theta$: 
\begin{align} 
    p(z_p|\theta) &= \frac{w_a(\theta)}{\sigma(\theta)} \left[ \frac{w_a(\theta_i)}{\sigma(\theta_i)}\right]^{-1}p(z_p|\theta_i) \; .
\end{align}
This equation can be understood as a vector-matrix-vector
multiplication
\begin{align} 
    p(z_p|\theta) &= \vec v(\theta) M^{-1}(\theta)\vec p(z_p)
    \quad \text{with} \quad 
    M_{ai} = \frac{w_a(\theta_i)}{\sigma(\theta_i)} \; .
    \label{eq:morphing_pz}
\end{align}
This morphing procedure does not only work for the parton-level
densities but also for the reco-level densities, following from
Eq.\eqref{eq:pxtheta_int}.  It can be exploited in different ways:
first, we can use it to generate training samples throughout the
$\theta$-space to train a neural network conditioned on $\theta$;
alternatively, we can perform morphing-aware likelihood estimation.

%%%%%%%%%%%%%%%%%%%%%%%%%%%%%%%%%%%%%%%%%%%%%%%%%%%%%%%%%%%%%%%%%%%
\subsubsection*{Morphing-aware likelihood estimation}

The morphing structure of Eq.\eqref{eq:morphing_pz} can be transferred
to the reco-level likelihood ratio
\begin{align}
    r(x|\theta,\theta_0) = \vec{v}(\theta)M^{-1} (\theta) \vec{r}(x) 
    \qquad \text{with} \quad 
    r_i(x) = r(x|\theta_i,\theta_0) \; ,
    \label{eq:morphing_r}
\end{align}
Using the BCE loss in Eq.\eqref{eq:BCE_ALICE_loss} and sampling only
from $\theta_i$ and $\theta_0$, we first train networks to approximate
$\vec r$.  These are then combined using Eq.\eqref{eq:morphing_r} to
obtain the complete likelihood ratio.  This way, the
$\theta$-dependence is not learned but directly imposed.

The morphing-aware likelihood-ratio estimation in \madminer works
slightly differently. Instead of learning the $r_i$ with individual
losses, the learned $\vec r_\varphi$ are combined into $r_\varphi$ for
which the MSE loss is calculated.  Consequently, the neural networks
have to be trained simultaneously using data covering $x$-space and
$\theta$-space. The sampling of $\theta$-space induces
instabilities~\cite{Brehmer:2018eca} and the morphing of the trained
networks for the likelihood ratio is an additional source of
uncertainty. Moreover, there is a high degeneracy in the scalar
product of $\vec v M^{-1}$ with $\vec r$ making it hard to regress on
$\vec r$. As we will demonstrate in Sec.~\ref{sec:toy}, learning the
likelihood ratio at every benchmark point leads to more stable
results, at least for low-dimensional problems. For higher-dimensional problems, a suitable choice of the morphing basis points becomes very important to ensure numerical stability.

While background contributions can in principle be handled as an additional contribution to $|\mat_\text{SM}(z_p)|^2$, it is advantageous for the network training to handle these separately via an additional classifier network. This is discussed in App.~\ref{app:backgrounds}.

%%%%%%%%%%%%%%%%%%%%%%%%%%%%%%%%%%%%%%%%%%%%%%%%%%%%%%%%%%%%%%%%%%%
\subsection{Derivative learning}
\label{sec:derivative_learning}

While morphing-aware likelihood learning relies on a set of benchmark
points, we can also use the structure of Eq.\eqref{eq:mat_factorize}.
Following Refs.~\cite{Chatterjee:2021nms,Chatterjee:2022oco}, we expand the unnormalized reco-level likelihood ratio
\begin{align}
   R(x|\theta,\theta_0) 
   \equiv \frac{d\sigma(x|\theta)/dx}{d\sigma(x|\theta_0)/dx} 
   = \frac{\sigma(\theta)p(x|\theta)}{\sigma(\theta_0)p(x|\theta_0)} \; .
\end{align}
around $\theta_0$ as
\begin{align}
       R(x|\theta,\theta_0) 
        = 1 + (\theta-\theta_0)_i R_i(x) 
            + (\theta-\theta_0)_i (\theta-\theta_0)_j R_{ij}(x) \; .
            \label{eq:Rexp}
\end{align}
The first and second derivatives
\begin{align}
    R_i(x) \equiv \frac{\partial}{\partial \theta_i} R(x|\theta,\theta_0)\bigg|_{\theta = \theta_0} 
    \qqquad\text{and}\qqquad
    R_{ij}(x) \equiv \frac{\partial^2}{\partial\theta_i \partial\theta_j} R(x|\theta,\theta_0)\bigg|_{\theta = \theta_0}
\end{align}
do not depend on $\theta$. Eq.~\eqref{eq:Rexp} is exact if
$R(x|\theta,\theta_0)$ is quadratic in $\theta$, as it is the case for
dimension-six SMEFT analyses.

As derived in Eq.\eqref{eq:lr_trick}, we can learn the derivatives of
$R(x|\theta,\theta_0)$ using a MSE loss. Using the parton-level joint
derivatives
\begin{alignat}{3}
    &f(z_p|\theta) \equiv f(z_p) = R_i(z_p) &&\equiv \frac{\partial}{\partial\theta_i}\frac{d\sigma(z_p|\theta)/dz_p}{d\sigma(z_p|\theta_0)/dz_p}\Bigg|_{\theta=\theta_0}
    &&= \frac{\partial_{\theta_i} |\mat(z_p|\theta)|^2}{|\mat(z_p|\theta_0)|^2}
    \Bigg|_{\theta_0} \notag \\
    &f(z_p|\theta) \equiv f(z_p) = R_{ij}(z_p) &&\equiv \frac{\partial^2}{\partial\theta_i\partial\theta_j}\frac{d\sigma(z_p|\theta)/dz_p}{d\sigma(z_p|\theta_0)/dz_p}\Bigg|_{\theta=\theta_0}
    &&= \frac{\partial_{\theta_i}\partial_{\theta_j} |\mat(z_p|\theta)|^2}{|\mat(z_p|\theta)|^2}
    \Bigg|_{\theta_0} \; ,
\end{alignat}
and sampling only from $\theta_0$, the learned functions will converge to
\begin{align}
 f_\varphi(x|\theta) \equiv R_{\varphi,i}(x) \approx R_i(x) 
 \qquad \text{and} \qquad 
 f_\varphi(x|\theta) \equiv R_{\varphi,ij}(x) \approx R_{ij}(x) \; .
\end{align}
The deduction proceeds in the same way as with the likelihood
ratio $r$, since the derivatives with respect to the theory parameters
can be pulled out of the integrals.

With this construction, we only need to simulate events at one point
in $\theta$-space, preferentially the SM point now assumed to be
$\theta_0 = 0$. Exploiting Eq.\eqref{eq:mat_factorize}, we then
calculate $R_i(z_p)$ and $R_{ij}(z_p)$, which can be obtained from the
simulator for the SM sample. After learning the reco-level
$R_{i,ij}(x)$, we obtain the likelihood ratio via
\begin{align}
    r(x|\theta,\theta_0) &=  \frac{\sigma(\theta_0)}{\sigma(\theta)}\frac{d\sigma(x|\theta)/dx}{d\sigma(x|\theta_0)/dx} = 
    \frac{\sigma(\theta_0)}{\sigma(\theta)} R(x|\theta,\theta_0).
\end{align}
The ratio of total cross-sections can be treated analogously to
Eq.\eqref{eq:Rexp},
\begin{align}
       \sigma(\theta)
        = \sigma(\theta_0) + (\theta-\theta_0)_i \sigma_i(\theta_0) 
            + (\theta-\theta_0)_i (\theta-\theta_0)_j \sigma_{ij}(\theta_0) \;,
\end{align}
with 
\begin{align}
    \sigma_i(\theta_0) \equiv \frac{\partial}{\partial \theta_i} \sigma(\theta)\bigg|_{\theta = \theta_0}
    \qqquad \text{and} \qquad
    \sigma_{ij}(\theta_0) \equiv \frac{\partial^2}{\partial\theta_i \partial\theta_j} \sigma(\theta)\bigg|_{\theta = \theta_0}\;.
\end{align}
These derivatives can be obtained by summing the parton-level
$R_{i,ij}(z_p)$ or the estimated $R_{\varphi,i}(x)$ and
$R_{\varphi,ij}(x)$ over the event sample. While the first option is
more precise, the second results in a more stable behaviour for small
$\theta$. This can be understood by expanding the summed
log-likelihood
\begin{align}
 \sum_i^N \log r(x_i|\theta,\theta_0)
 &\simeq \left[ \sum_i^N R_{\varphi,j}(x_i) -  N \frac{\sigma_j}{\sigma(\theta_0)}\right] 
 \; (\theta_j - \theta_{0,j}) + \order\left((\theta_j - \theta_{0,j})^2\right) = \notag\\
 &= \order\left((\theta_j - \theta_{0,j})^2\right)\;.
\end{align}
If the $\sigma_j$ are calculated by summing $R_{j}(z_p)$ and
$R_{\varphi,j}(x)$ are not learned perfectly, the first-order term
will not completely vanish for $\theta$ close to
$\theta_0$. Calculating the derivatives by summing $R_{\varphi,i}(x)$
and $R_{\varphi,ij}(x)$ enforces a vanishing first-order term by
construction.

It is important to note that this derivative-learning method fails if
the weights change significantly within the considered $\theta$-range
in some $x$-region. In that case a region with very large weights will
be undersampled, leading to numerical instabilities in the most
sensitive phase space regions.

%%%%%%%%%%%%%%%%%%%%%%%%%%%%%%%%%%%%%%%%%%%%%%%%%%%%%%%%%%%%%%%%%%%
\subsection{Fractional smearing}

When simulating reco-level events $x$ from the parton level $z_p$, all
4-momenta are smeared out by parton shower and detector effects. A
unit-weight event at the parton level turns into a unit-weight event
at the reco-level. To estimate the changing phase-space densities in
sparsely populated regions more efficiently, we can describe the
smearing using a set of weighted events,
given that the network training can trivially be extended to this case~\cite{Backes:2020vka}.

To this end, we pass some events through the simulation chain 
multiple times, for instance events with large parton-level likelihood 
ratio or with large parton-level derivatives.
These events contribute to the loss function with a reduced 
weight, to ensure that the event sample still follows the distribution 
$p(x|z_p)p(z_p|\theta)$. 

Let us assume that we want to learn the target function $r(z_p|\theta,\theta_0)$, 
omitting the index in $\theta_i$.
We follow a series of steps manipulating the parton-level events:
\begin{enumerate}
    \item calculate the mean $\mu$ and standard deviation $\sigma$ of the target $r(z_p|\theta,\theta_0)$;
    \item assign a fractional-smearing weight $w$ to every event, initially set to one;
    \item smear an event by copying it $n$ times and assigning $w=1/n$ to each copy;
    \item define a threshold, for instance smear all events with $|w \times r(z_p|\theta,\theta_0) -\mu| > t \sigma$ for a given $t$;
    \item smear until the threshold value is reached.
\end{enumerate}
This sample of smeared and weighted parton level events is passed through the simulation. The fractional-smearing 
weights are incorporated into the MSE loss, for example for the likelihood ratio regression for a fixed $\theta$,
\begin{align}
    \loss 
    = \XXLangle \left[ r(z_p|\theta,\theta_0)- r_\varphi(x|\theta) \right]^2 \XXRangle_{x,z_p \sim p(x|z_p)p(z_p|\theta)} 
    = \sum_i^N \; w_i \left[ r(z_{p,i}|\theta,\theta_0)- r_\varphi(x_i|\theta) \right]^2\;,
\label{eq:loss_fs}
\end{align}
illustrating the role of the weights $w$. The sum runs over the event sample. If we apply cuts to the reco-level events, only 
some of the fractionally-smeared events belonging to the same parton-level event might survive.
This is not a problem, since the cut event sample still corresponds to a weighted sample of 
the distribution $p(x|z_p)p(z_p|\theta)$.

As an alternative to using fractional smearing for preparing the neural-network training dataset, boosted-decision trees which learn to ``bin'' the phase space can be used. As has been demonstrated in Refs.~\cite{Chatterjee:2021nms,Chatterjee:2022oco} for the derivative learning approach, the division of the phase space allows for accurate estimates even in sparsely populated phase-space regions.

%%%%%%%%%%%%%%%%%%%%%%%%%%%%%%%%%%%%%%%%%%%%%%%%%%%%%%%%%%%%%%%%%%%
\subsection{L-GATr}

To improve the likelihood learning we can employ equivariant neural networks, which have been shown to enhance the performance for different problems in particle physics~\cite{Gong:2022lye,Ruhe:2023rqc,Bogatskiy:2023nnw,Spinner:2024hjm}. These models encode knowledge about the spacetime symmetry directly into their structure. This way, the network does not need to spend training resources in learning the symmetry properties of the data, and its operations get restricted to those allowed by the symmetry. This makes equivariant networks fast to train, resistant to overfitting, and sample-efficient.

We use the Lorentz-Equivariant Geometric Algebra Transformer (L-GATr)~\cite{Spinner:2024hjm,brehmer2023geometric,lgatr} for likelihood learning. L-GATr is a transformer-based model that processes data in the spacetime geometric algebra representation.
To construct it, a geometric product is introduced to the vector space to  create higher-order objects from the original Lorentz-vectors. This operation modifies the properties of the basis of the vector space starting from the relation
\begin{align}
    \left\{\gamma^{\mu}, \gamma^{\nu} \right\} = 2 g^{\mu\nu} \; .
    \label{eq:gamma-matrices}
\end{align}
This anti-commutation relation defines gamma matrices and establishes a close connection between the real spacetime algebra and the complex Dirac algebra. Inspired by this connection, we build multivectors forming the spacetime algebra:
\begin{align}
    x = x^S \; 1 
    + x^V_\mu  \; \gamma^\mu 
    + x^B_{\mu\nu} \; \sigma^{\mu\nu} 
    + x^A_\mu \; \gamma^\mu \gamma^5 
    + x^P \; \gamma^5 
    \qquad \text{with} \qquad 
    \begin{pmatrix}
    x^S \\ x^V_\mu \\ x^B_{\mu\nu} \\ x^A_\mu \\ x^P 
    \end{pmatrix} 
    \in\mathbb{R}^{16} \; .
\label{eq:multivector}
\end{align}
A multivector consists of 16 components organized in grades according to the length of  $\gamma^\mu$-matrix products needed to express them. Specifically, $x^S 1$ represents a scalar, $x^V_\mu \gamma^\mu$ a vector, $x^B_{\mu\nu} \sigma^{\mu\nu}$ a geometric bilinear, $x^A_\mu \gamma^\mu \gamma^5$ an axial vector, and $x^P \gamma^5$ a pseudoscalar. 

The spacetime algebra not only expresses a wide range of objects in Minkowski space, it also offers a way to define learnable equivariant transformations on particle physics data. A transformation $f$ is equivariant with respect to Lorentz transformations $\Lambda$ if
\begin{align}
    f\Big(\Lambda(x)\Big) = \Lambda \Big(f(x)\Big) \; .
\end{align}
We encode this condition in L-GATr by considering the fact that every grade transforms under a given representation of the Lorentz group. This structure of the geometric algebra allows us to easily develop equivariant versions of any network architecture. L-GATr performs this adaptation using transformer structures, with equivariant versions of linear, attention, layer-normalization, and activation operations,
\begin{align}
    \bar x &= \text{LayerNorm}(x) \notag \\
    \text{AttentionBlock}(x) &= \text{Linear}\circ \text{Attention}(\text{Linear}(\bar x), \text{Linear}(\bar x), \text{Linear}(\bar x)) + x \notag \\
    \text{MLPBlock}(x) &= \text{Linear}\circ \text{Activation}\circ \text{Linear}\circ \text{GP}(\text{Linear}(\bar x), \text{Linear}(\bar x))+x \notag \\
    \text{Block}(x) &= \text{MLPBlock}\circ \text{AttentionBlock}(x) \notag \\
    \text{L-GATr}(x) &= \text{Linear}\circ \text{Block}\circ \text{Block}\circ\cdots \circ \text{Block}\circ\text{Linear}(x) \; .
\end{align}
We also include the MLPBlock containing a geometric product operation to maximize network expressivity. Further details are provided in Refs.~\cite{Spinner:2024hjm, brehmer2023geometric,lgatr}.

For likelihood learning our data consists of particle properties organized as tokens. For the L-GATr input we embed the 4-momenta $p^\mu$ as the vector grade in the algebra,
\begin{align}
    x^V_{\mu} = p_{\mu}
    \qquad \text{and} \qquad 
    x^S = x^T_{\mu\nu}=x^A_\mu=x^P=0 \; . 
    \label{eq:embedding}
\end{align}
In addition to this geometric vector, our data also contains particle identification and partial kinematic input, like missing transverse momentum. They are given to the network as independent scalar inputs and are evaluated by the transformer in parallel to the multivectors. Both tracks are mixed in the equivariant linear layers. 
As L-GATr output we select the scalar component of the corresponding multivectors. In our case, we perform this operation on a global token, an extra particle object that is appended to the rest of the particle tokens in each sample. This extra token is empty at the input level and gains meaning at the output level through the network training~\cite{lgatr}. 

%%%%%%%%%%%%%%%%%%%%%%%%%%%%%%%%%%%%%%%%%%%%%%%%%%%%%%%%%%%%%%%%%%%
\section{Toy model}
\label{sec:toy}

To illustrate the ideas behind our SBI tools we use a toy model with two Gaussians 
at parton level,
\begin{align}
    p_{\alpha}(z_p|\theta) = 
    \frac{\normal_{0,1} (z_p) + \theta^2 \normal_{\alpha,0.1}(z_p)}{1+\theta^2} \; ,
\end{align}
where $\theta$ stands for the theory parameter we want to constrain. The parameter $\alpha$ 
controls the distance between the two Gaussians.
We then add a simple Gaussian smearing,
\begin{align}
    p(x|z_p) = \normal_{0,0.7}(z_p) \; ,
\end{align}
emulating parton shower and detector. $x$ represents a reco-level observable. For this 
toy model, the reco-level likelihood becomes
\begin{align}
    p_{\alpha}(x|\theta) = 
    \frac{\normal_{0,1.22}(x) + \theta^2 \normal_{\alpha,0.71}(x)}{1+\theta^2} \; .
\end{align}
Then, the likelihood ratio is given by
\begin{align}
    r_{\alpha}(x|\theta,\theta_0) = 
    \frac{1 + \theta_0^2}{1+ \theta^2}\frac{\normal_{0,1.22}(x) + \theta^2 \normal_{\alpha,0.71}(x)}{\normal_{0,1.22}(x) + \theta_0^2 \normal_{\alpha,0.71}(x)} \; .
\end{align}
In addition to the reduced dimensionality, the main 
simplifications of this toy model is that we know the normalization exactly and that we 
do not consider any continuum backgrounds.

%%%%%%%%%%%%%%%%%%%%%%%%%%%%%%%%%%%%%%%%%%%%%%%%%%%%%%%%%%%%%%%%%%%
\subsection{Morphing-aware  vs.~derivative learning}
\label{sec:toy_compare}

%--------------------------------
\begin{figure}[t]
    \includegraphics[width=0.495\textwidth,trim={.5cm 0cm .5cm 0cm},clip]{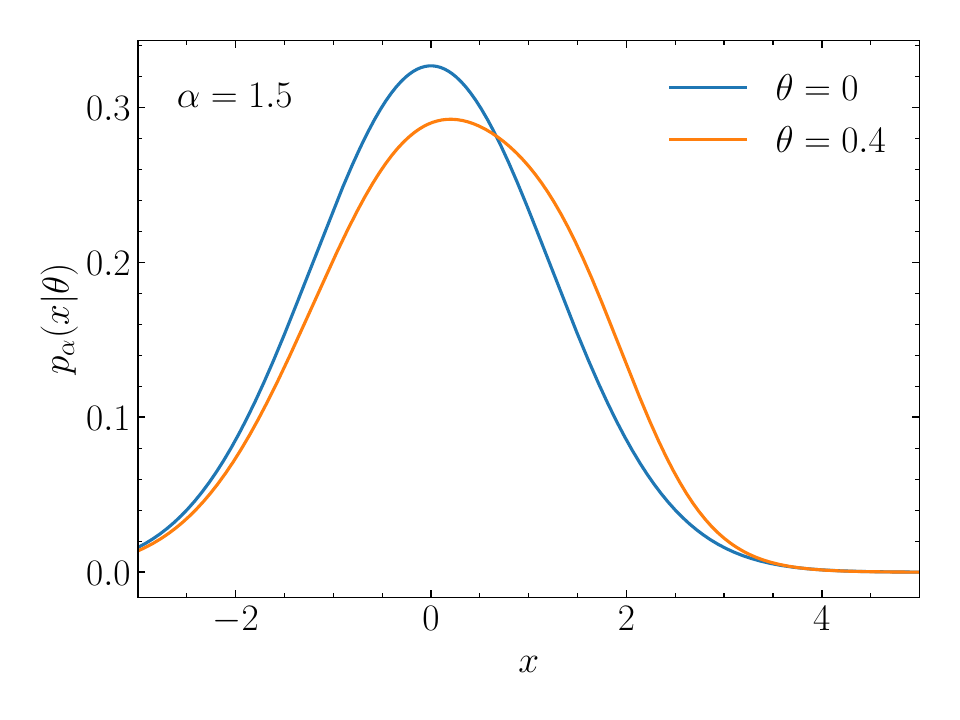}
    \includegraphics[width=0.495\textwidth,trim={.5cm 0cm .5cm 0cm},clip]{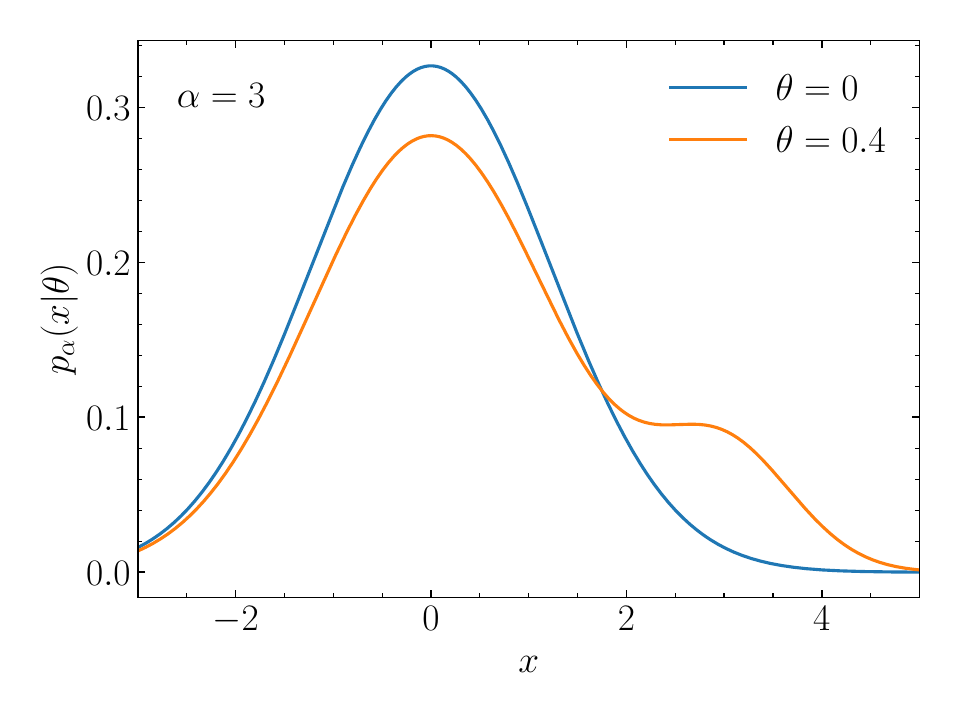}
    \caption{Likelihoods for local and non-local cases. Both show the full $p(x|\theta)$ for $\theta=0$ and $\theta=0.4$.}
    \label{fig:toy_mod_signal}
\end{figure}
%--------------------------------

First, the Gaussian toy model allows us to compare morphing-aware and 
derivative likelihood estimation. The locality in phase space is determined by $\alpha$. For $\alpha\sim 1$ --- i.e., if it is similar to the widths of the Gaussians ---, the likelihood 
will hardly vary when we change $\theta$ away from $\theta=0$. On the 
other hand, for $\alpha \gg 1$ even small changes in $\theta$ will shift the 
bulk of the likelihood, as illustrated 
for $\alpha=1.5$ and $\alpha=3$ in Fig.~\ref{fig:toy_mod_signal}.
This means that for $\alpha\sim 1$ we can sample from the 
$\theta=0$-hypothesis, while for $\alpha \gg 1$ we need to sample from 
different basis points using a morphing-aware approach.

To test derivative learning, we generate 
$3\times10^5$ events at the reference point $\theta_0 = 0$. 
For morphing-aware sampling, we generate $10^5$ events at
$\theta = -1, 0, 1$. For both, we use the same number 
of networks with the same amount of parameters and training epochs. 
Each network is implemented as a repulsive ensemble~\cite{DBLP:journals/corr/abs-2106-11642,Rover:2024pvr,Plehn:2022ftl},
to estimate the uncertainty due to limited training data. For testing 
the likelihood ratio estimates, we use data generated with  
$\theta = 0.6$.

%--------------------------------
\begin{figure}[b!]
    \includegraphics[width=0.495\textwidth,trim={.5cm 0cm .5cm 0cm},clip]{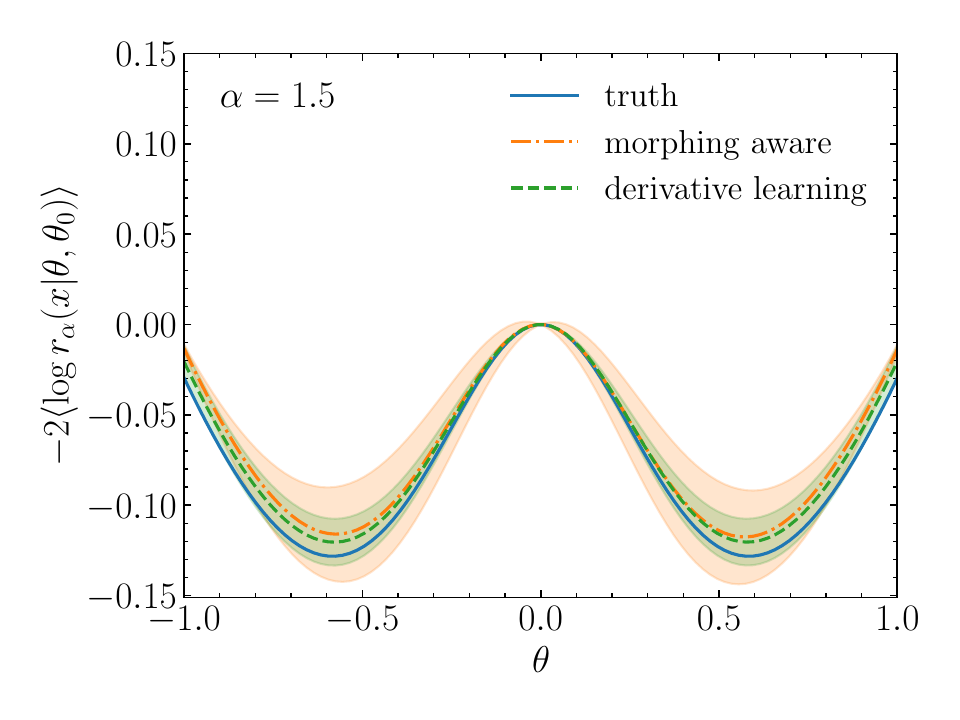}
    \includegraphics[width=0.495\textwidth,trim={.5cm 0cm .5cm 0cm},clip]{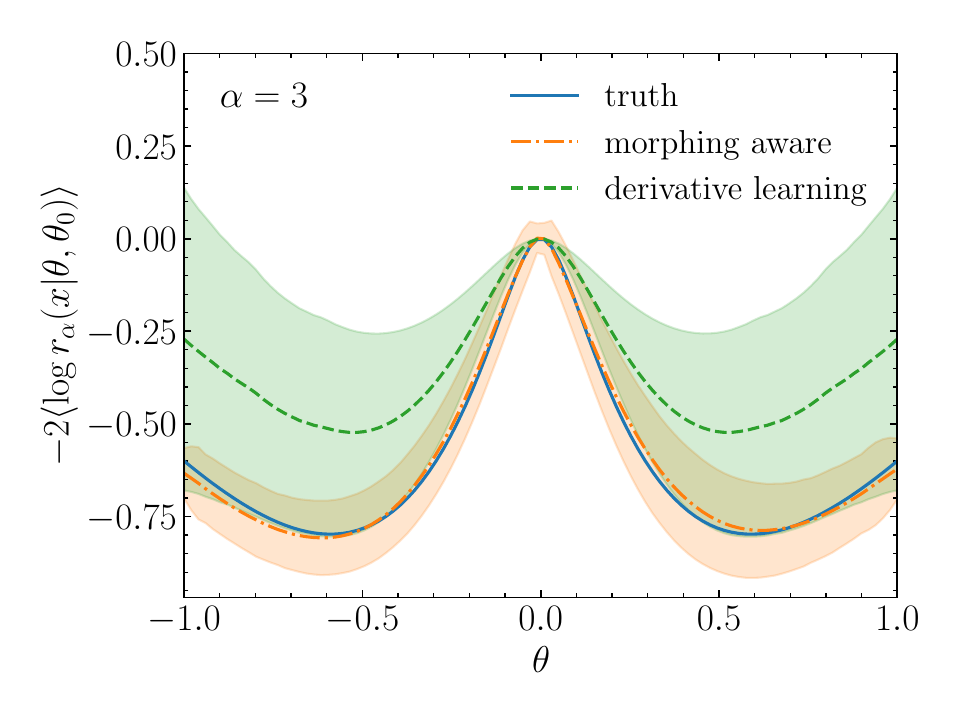}\\
    \caption{Comparison between the likelihood ratios estimated using the morphing-aware and derivative learning approaches for $\alpha=1.5$ (left) and $\alpha=3$ (right). The truth value is $\theta = 0.6$.}
    \label{fig:toy_comparison}
\end{figure}
%--------------------------------

The results of the toy model reflect two aspects of the likelihood 
ratio learning problem. In the left panel of 
Fig.~\ref{fig:toy_comparison} we see that for $\alpha=1.5$ the two methods 
have comparable performance and can be trusted.
For $\alpha=3$, the derivative learning method
gives worse results. The training data at the $\theta = 0$ hypothesis
does not cover the relevant phase-space regions, as we can already 
see in Fig.~\ref{fig:toy_mod_signal}. Finally, the 
repulsive ensembles give us an indication 
on how reliable each method is, for instance 
comparing the uncertainty bands for $\alpha=3$.

%--------------------------------
\begin{figure}[t]
    \centering
    \includegraphics[width=0.5\textwidth,trim={.5cm 0cm .5cm 0cm},clip]{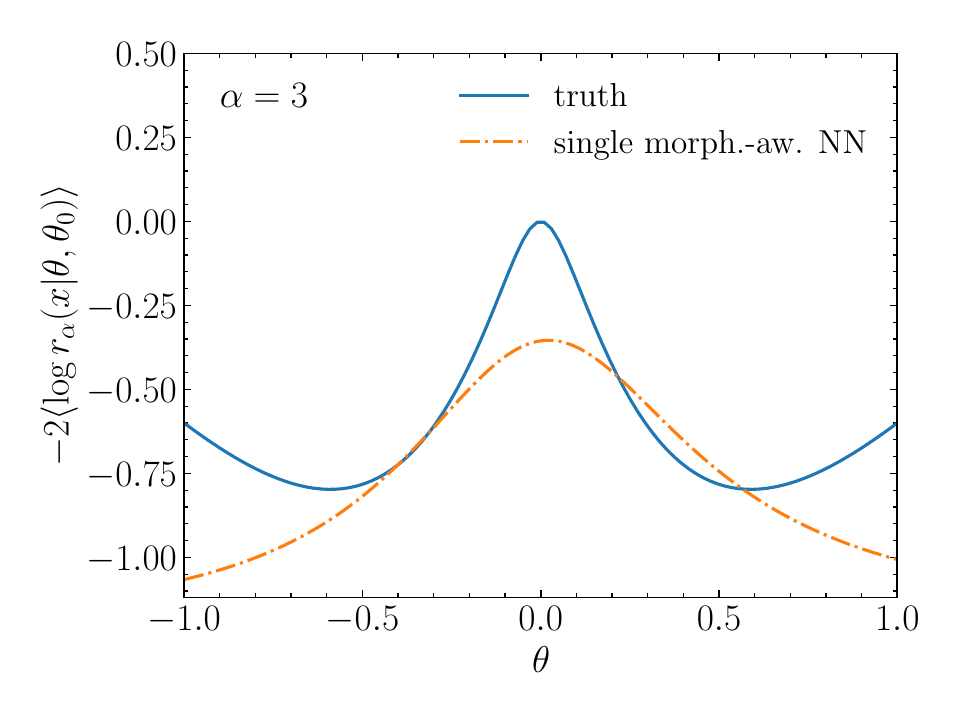}\\
    \caption{Likelihood ratio estimated using the single neural network morphing-aware 
    implementation of Ref.~\cite{Brehmer:2018eca}, compared to 
    the truth.}
    \label{fig:madminer_morph}
\end{figure}
%--------------------------------

For our toy model, morphing-aware sampling~\cite{Brehmer:2018eca} 
provides an accurate likelihood ratio estimation. 
As explained in Sec.~\ref{sec:morphing}, the novelty in 
our implementation is that we train the likelihood ratios for each 
basis point independently. We can compare our implementation 
with the version from Ref.~\cite{Brehmer:2018eca}, again with 
the same size of the training dataset and the same number of network 
parameters and training epochs. For the $\theta$-prior in 
Eq.\eqref{eq:MSE_ratio_loss}, we use a standard Gaussian.
We only check the non-local case, because this is where
morphing-aware sampling has advantages. The results are shown in 
Fig.~\ref{fig:madminer_morph}, indicating that for the toy model
our implementation is more stable.

The stability of our morphing-aware likelihood estimation can, however, be compromised for more than one theory parameter. In this case, the choice of basis points becomes very important, and a non-optimal choice will lead to large coefficients in Eq.\eqref{eq:morphing_r} from the matrix inversion. This, in turn, results in enhanced uncertainties due to imperfect network training.

%%%%%%%%%%%%%%%%%%%%%%%%%%%%%%%%%%%%%%%%%%%%%%%%%%%%%%%%%%%%%%%%%%%
\subsection{Fractional smearing}
\label{sec:toy_frac}

To illustrate our novel fractional smearing, we use the same toy model 
as before in an even less local setting with $\alpha = 4$. This models 
a SMEFT situation, where kinematic tails can be particularly sensitive to 
the theory parameters. 
We use the more challenging derivative learning and generate 
$10^5$ training events for $\theta = 0$. This way, the 
large-$x$ region, which has the largest sensitivity to $\theta$
and a large derivative $R_{\theta^2}$, is only sparsely populated. 

%--------------------------------
\begin{figure}[b!]
    \centering
    \includegraphics[width=0.5\textwidth]{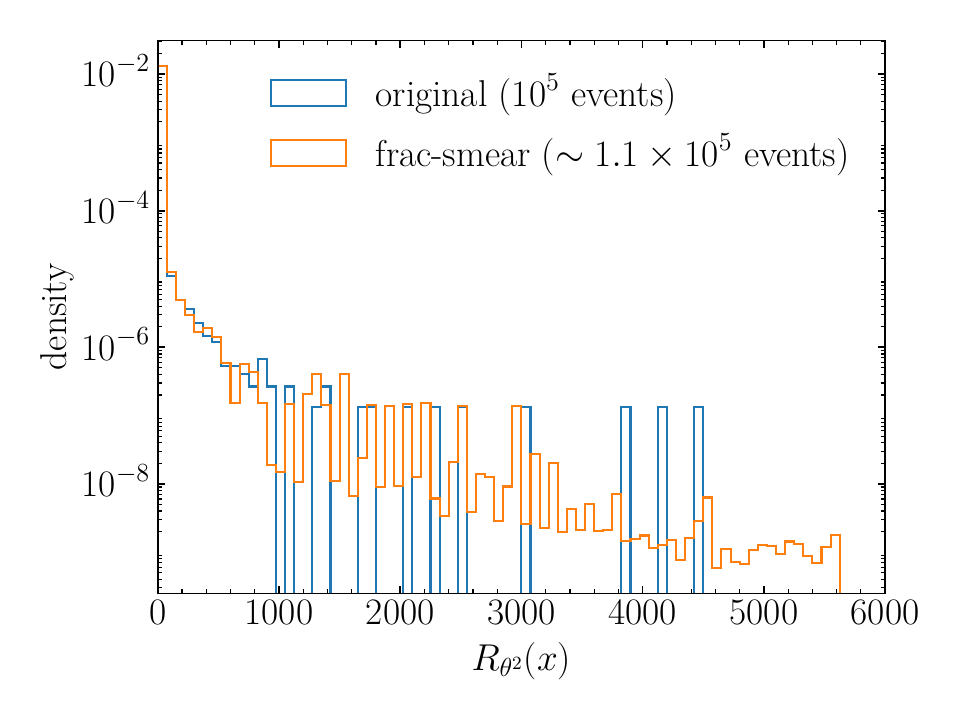}
    \caption{Histogram of the reco-level second derivative $R_{\theta^2}$, for the original dataset (blue) and the 
    fractionally-smeared dataset (orange).}
    \label{fig:fractional_smearing_hist_der_x_2}
\end{figure}
%--------------------------------

%--------------------------------
\begin{figure}[t]
    \includegraphics[width=0.49\textwidth,trim={.5cm 0cm .5cm 0cm},clip]{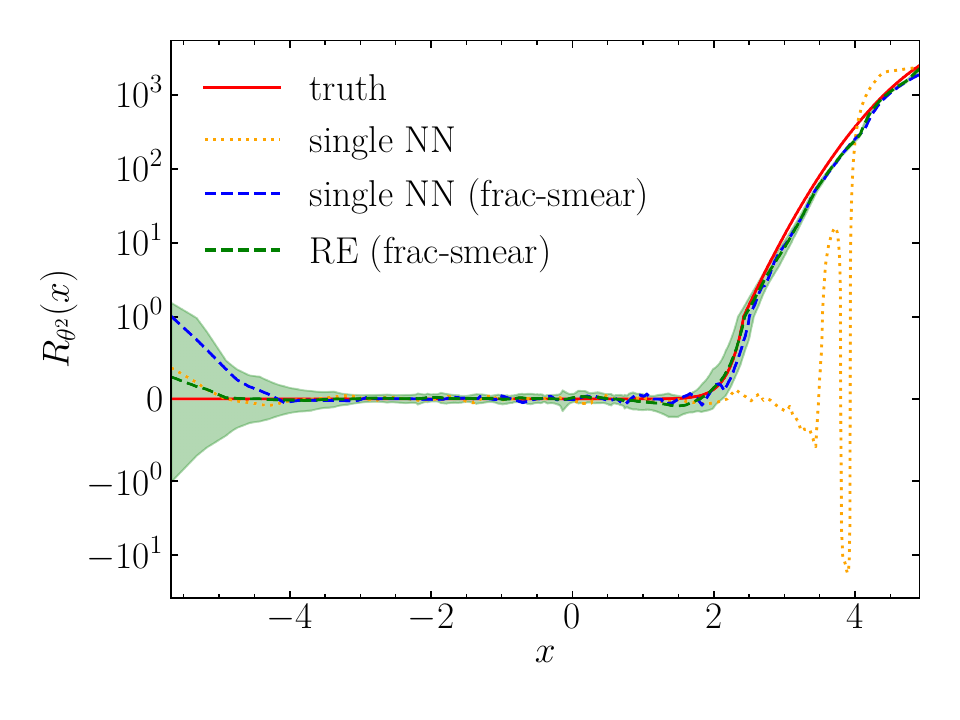}
    \includegraphics[width=0.49\textwidth,trim={.5cm 0cm .5cm 0cm},clip]{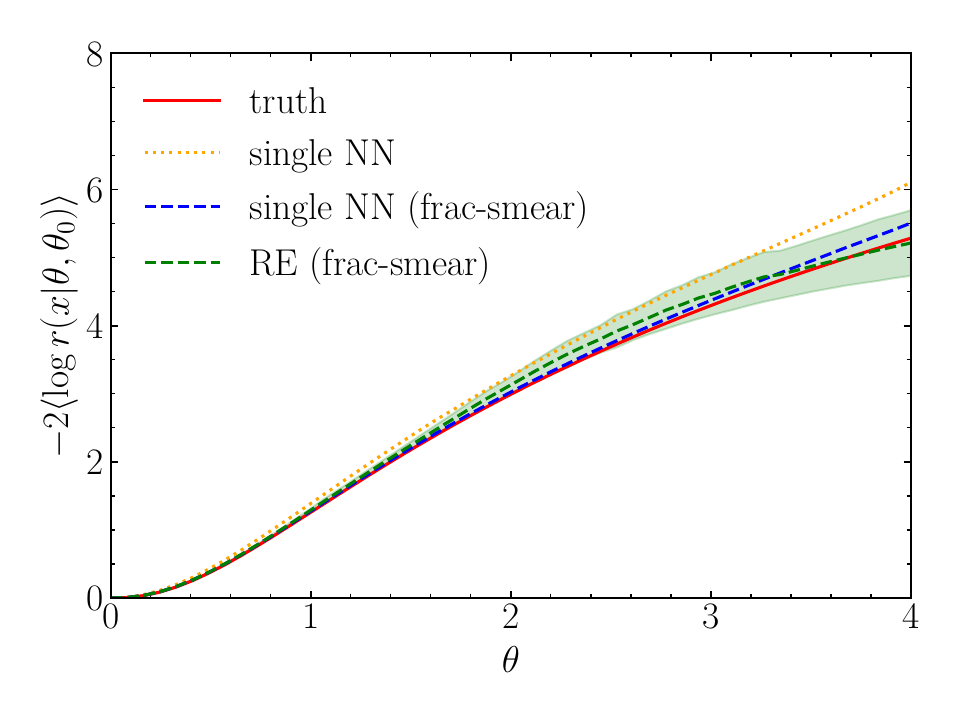}
    \caption{Left: learned second derivative as a function $x$ for various setups compared to the truth. Right: expectation value of the learned likelihood ratio as a function of $\theta$ compared to the true likelihood ratio.}
    \label{fig:fractional_smearing_results}
\end{figure}
%--------------------------------

Figure~\ref{fig:fractional_smearing_hist_der_x_2} shows the 
distribution of the reco-level second derivative with respect 
to $\theta$, for the original and fractionally-smeared datasets. The 
original dataset only features a few isolated events with large 
$R_{\theta^2}$. The underlying distribution is hard to learn.
The fractionally-smeared dataset has a much more balanced 
distribution, which is far easier to learn. Additional 
visualizations of the datasets are provided in App.~\ref{sec:frac_smearing_datasets}.

Next, we show the learned second derivatives as a function of $x$ 
in the left panel of Fig.~\ref{fig:fractional_smearing_results}. 
The single network and the repulsive ensemble trained on the 
fractional-smearing dataset capture the large-$x$ behaviour of the 
truth much better than the neural network trained on the original 
dataset. Finally, we show the expected log-likelihood ratios 
in the right panel of Fig.~\ref{fig:fractional_smearing_results}. 
Again, the networks trained on the fractionally-smeared dataset are 
much closer to the true likelihood ratio, illustrating the 
significantly improved training.

%%%%%%%%%%%%%%%%%%%%%%%%%%%%%%%%%%%%%%%%%%%%%%%%%%%%%%%%%%%%%%%%%%%
\section{LHC application: \texorpdfstring{$pp\to W^\pm Z$}{pp->WZ}}
\label{sec:lhc}

%--------------------------------
\begin{figure}[t]
    \centering
    \includegraphics[width=0.55\textwidth]{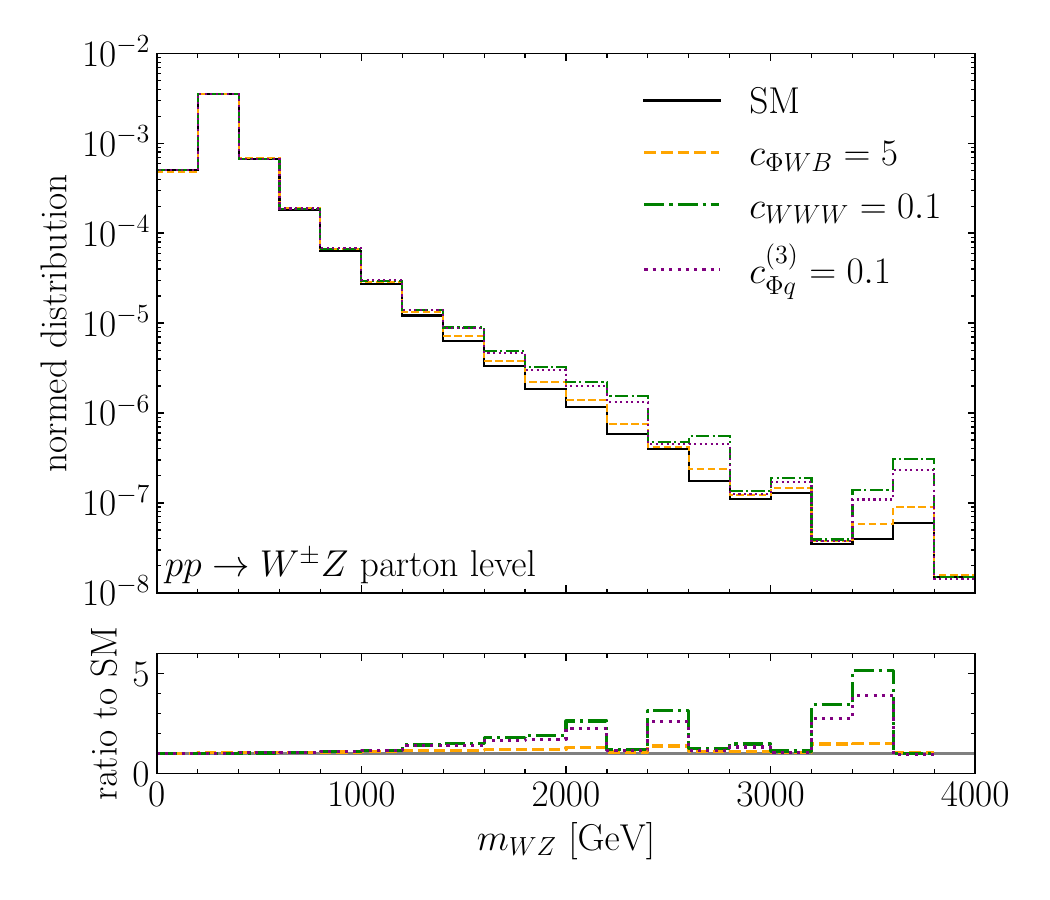}
    \caption{Normalized distribution of the $WZ$ invariant mass for
      the SM and three Wilson coefficients, one at a time.}
    \label{fig:WZ_mWZ}
\end{figure}
%--------------------------------

As an LHC physics example, we use $W^\pm Z$ production at Run~3, with
$\sqrt{s}=13.6\,\tev$ and $\mathcal{L}=300\,\ifb$. At tree level,
this process has two contributions, distinguished by the $Z$-coupling
either to quarks or to the $W$-boson. We modify the SM-interactions
through the dimension-6 operators
\begin{align}
    \mathcal{O}_{\Phi WB} 
    &= \Phi^\dagger\tau^a\Phi W^a_{\mu\nu}B^{\mu\nu} \notag \\
    \mathcal{O}_{WWW} &= \epsilon^{abc} W_\mu^{a\nu}W_\nu^{b\rho}W_\rho^{c\mu} \notag \\
    \mathcal{O}_{\Phi q}^{(3)} &= (\Phi^\dagger i \overset{\leftrightarrow}{D^a}_\mu \Phi)(\bar Q_L\tau^a \gamma^\mu Q_L) \;.
\end{align}
Here, $\Phi$ is the Higgs doublet, $W^{\mu\nu}$ and $B^{\mu\nu}$ are
the $SU(2)_L$ and $U(1)_Y$ field strengths, $D_\mu$ is the covariant
derivative, and $Q_L$ the left-handed quark doublets.  The UV cutoff
$\Lambda$ is set to $1~\tev$ for our numerical analysis. In the $WZ$ process, the first
operator modifies the $Z$-couplings to quarks and the
$WWZ$-coupling. The second modifies only the
$WWZ$-coupling; the third one, only the $Z$ coupling to quarks. The corresponding Wilson coefficients are our theory
parameters,
\begin{align}
  \theta = \left( c_{\Phi WB}, c_{WWW}, c_{\Phi q}^{(3)} \right)
  \qquad \text{with} \qquad
  \theta_0 = (0,0,0) \; .
\end{align}
We use \madgraph~3.5.0~\cite{Alwall:2011uj} for event generation at
the leading order, employing the
\textsc{SMEFTatNLO}~\cite{Degrande:2020evl} \textsc{UFO} model. For the
reco-level analysis, we decay the vector bosons leptonically
using \madspin~\cite{Artoisenet:2012st}. The parton shower is
\pythia~8.306~\cite{Sjostrand:2014zea}, the detector simulation
\delphes~3.5.0~\cite{deFavereau:2013fsa}, and the jet algorithm
\fastjet~3.3.4~\cite{Cacciari:2011ma}.

The NLO QCD corrections have a sizeable dependence on the phase space region and the entering Wilson coefficients~\cite{Baglio:2019uty}. This dependence is, however, reduced by applying a jet veto --- as done below for our reco-level analysis. Therefore, we approximate the NLO QCD corrections by multiplying the
total rate with an NLO $K$-factor~\cite{Denner:2020eck}.

We illustrate the effect of each Wilson coefficient on the
distribution of the $WZ$-invariant mass in
Fig.~\ref{fig:WZ_mWZ}. The effects of $c_{WWW}$ and
$c_{\Phi q}^{(3)}$ on the high-energy tail are specially visible.

%%%%%%%%%%%%%%%%%%%%%%%%%%%%%%%%%%%%%%%%%%%%%%%%%%%%%%%%%%%%%%%%%%%
\subsection{Parton level}

%--------------------------------
\begin{figure}[b!]
    \includegraphics[width=\textwidth,trim={3.cm 0cm 4cm 0cm},clip]{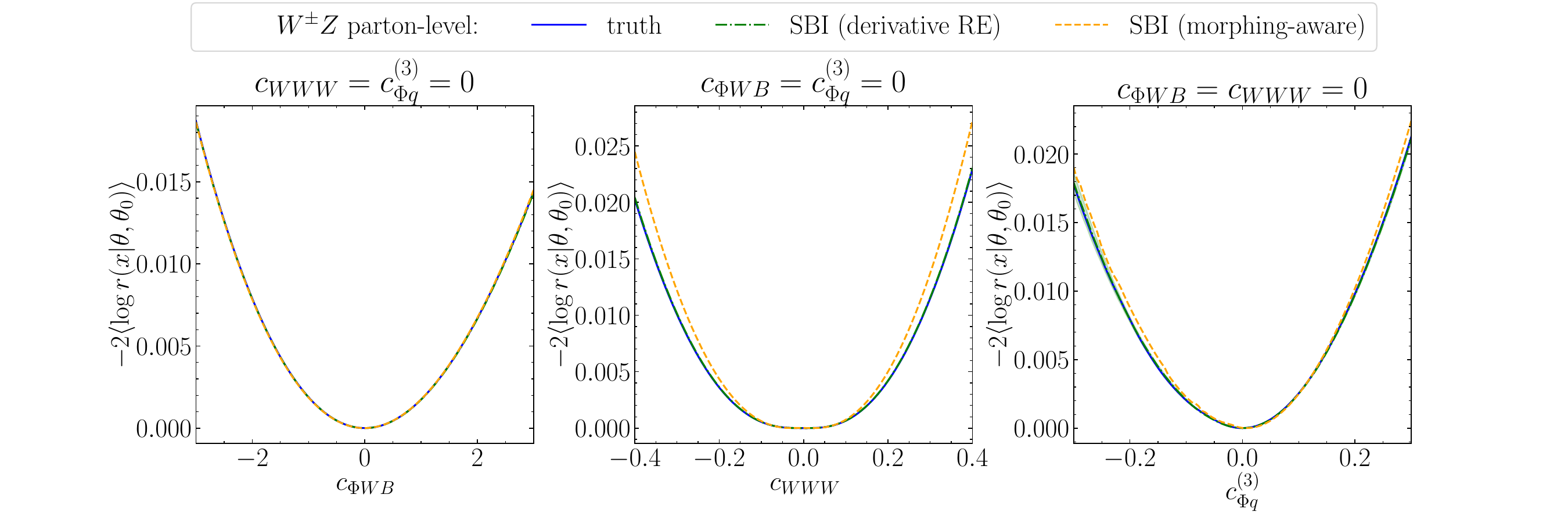}
    \caption{Log-likelihood ratio averaged over the event sample,
      showing the truth, the morphing-aware SBI result, and the derivative learning SBI result using repulsive ensembles as a function of $c_{\Phi WB}$ (left), of
      $c_{WWW}$ (center), and of $c_{\Phi q}^{(3)}$ (right).}
    \label{fig:WZ_parton_1D}
\end{figure}
%--------------------------------

At parton level, we apply derivative learning at the SM point to learn the likelihood ratio employing repulsive ensembles. We tested that at parton level L-GATr yields only marginal improvements. This is due to the simplicity of the task, there is not much to be gained from increasing the sophistication of the network. Moreover, we also show results using the morphing-aware approach.

We compute the parton-level derivatives and likelihood ratios using the built-in reweighting functionality of \madgraph. It is important to reweight to the full
amplitude and not only the amplitude corresponding to the
helicity of each event, as it is done by default in \madgraph\ --- see
Refs.~\cite{Mattelaer:2016gcx,Belvedere:2024nzh} for more details.

%--------------------------------
\begin{figure}[t]
    \includegraphics[width=\textwidth,trim={3.5cm 0cm 4cm 0cm},clip]{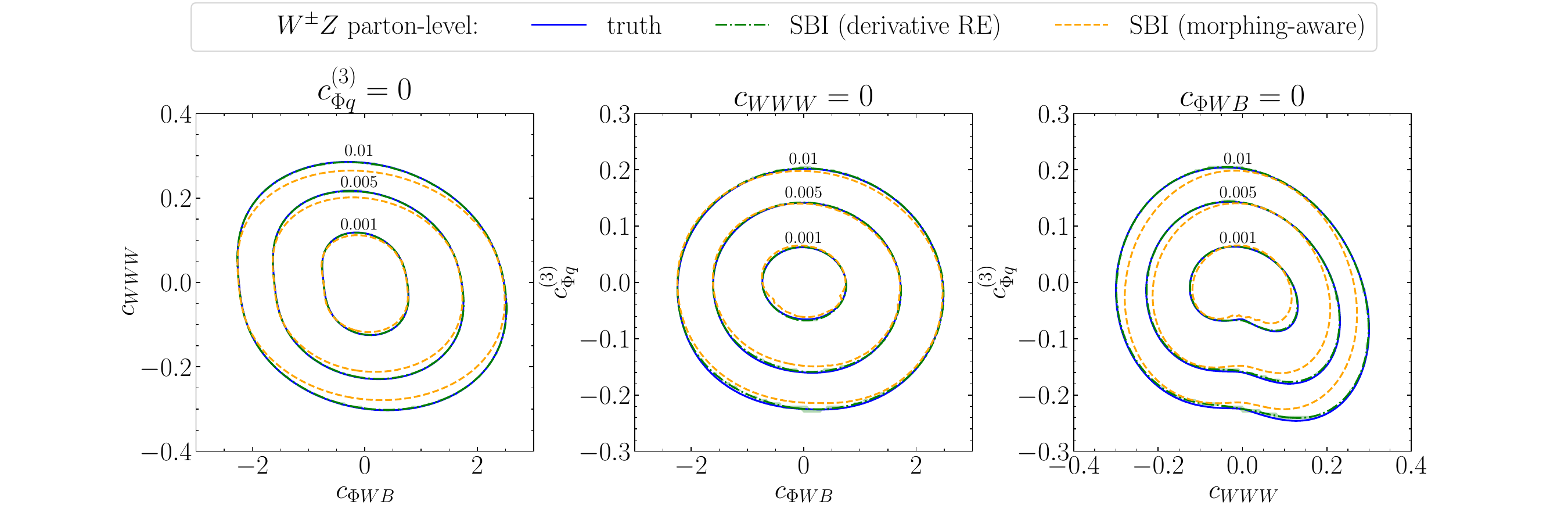}
    \caption{Contours of the log-likelihood ratio averaged over the
      event sample showing the truth, the morphing-aware SBI result, and the derivative learning SBI
      result using repulsive ensembles as a function of $c_{\Phi WB}$
      and $c_{WWW}$ (left), of $c_{\Phi WB}$ and $c_{\Phi q}^{(3)}$
      (center), and of $c_{WWW}$ and $c_{\Phi q}^{(3)}$ (right).}
    \label{fig:WZ_parton_2D}
\end{figure}
%--------------------------------

%--------------------------------
\begin{figure}[b!]
    \includegraphics[width=\textwidth,trim={3.5cm 0cm 4cm 0cm},clip]{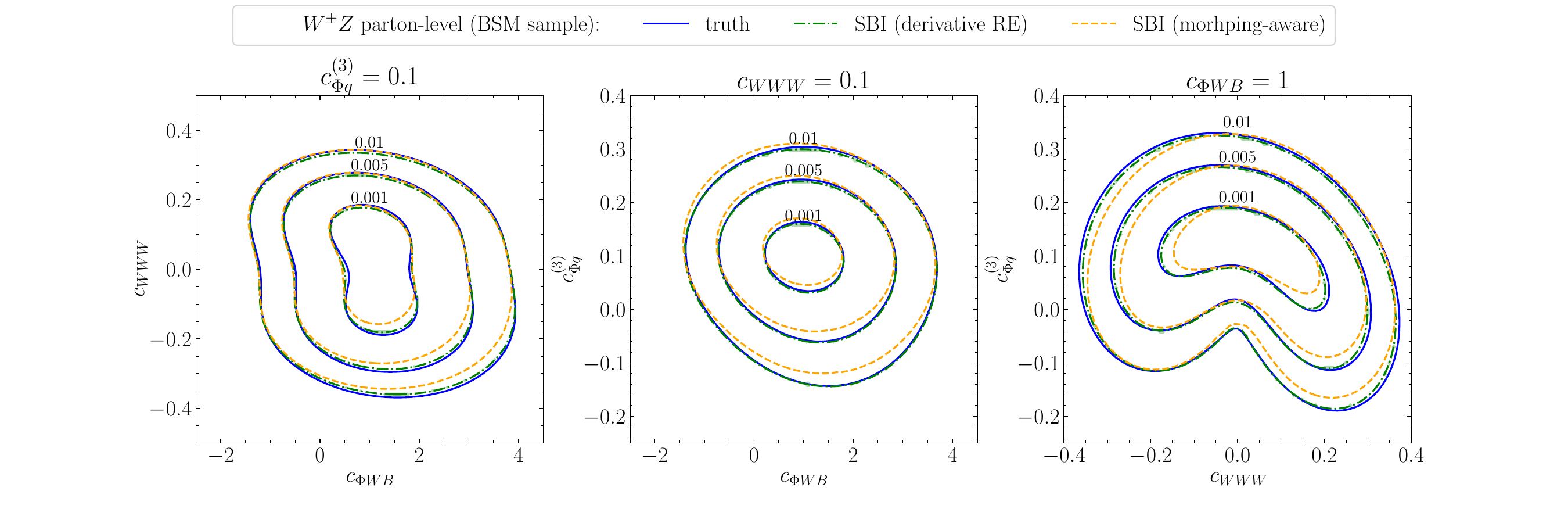}
    \caption{Contours of the log-likelihood ratio averaged over the
      BSM event sample showing the truth, the morphing-aware SBI result, and the derivative learning SBI
      result using repulsive ensembles as a function of $c_{\Phi WB}$
      and $c_{WWW}$ (left), of $c_{\Phi WB}$ and $c_{\Phi q}^{(3)}$
      (center), and of $c_{WWW}$ and $c_{\Phi q}^{(3)}$ (right).}
    \label{fig:WZ_parton_2D_BSM}
\end{figure}
%--------------------------------

As input features for the parton-level training, we use the Mandelstam
variables $s$ and $t$, together with the charge of the final-state
$W$. We evaluate the likelihood ratio using data samples generated at the
SM point and at the exemplary BSM point $c_{\Phi WB} = 1, c_{WWW} =
c_{\Phi q}^{(3)} = 0.1$. We train the derivative networks at the SM point $\theta_0$ using a training dataset containing 300k
events and a validation dataset using 100k events. For morphing-aware learning, each training dataset is composed of 250k SM points and 250k BSM points. The choice of morphing basis points is detailed in App.~\ref{app:morphing-aware LHC}. The SM and BSM test
datasets contain 100k and 300k events, respectively. For derivative learning, we scale the input as well as output
features to mean zero and standard deviation one.

First, we look at the average log-likelihood ratio using the SM
sample. We show the results of the derivative learning and the morphing-aware approach for a single
non-zero Wilson coefficient in Fig.~\ref{fig:WZ_parton_1D}. For the derivative learning approach, the
differences to the true log-likelihood ratio are orders of magnitude
smaller than the overall log-likelihood, and the uncertainty estimate
using repulsive ensembles is correspondingly small. This reflects that
the training dataset and the evaluation dataset are generated at the
same $\theta$-point. The morhping-aware approach performs worse but still provides a good estimation of the true likelihood ratio. The largest deviations from the true log-likelihood ratio appear across the $c_{WWW}$ dimension. A possible explanation --- as can be inferred by the flat shape close to zero of the true curve --- could be that the quadratic SMEFT operator insertion dominates over the linear one, complicating the choice of suitable morphing basis points. 

Similarly, the 2-dimensional correlations from
derivative learning, displayed in Fig.~\ref{fig:WZ_parton_2D}, are in
excellent agreement with the true log-likelihood ratio. The
derivative learning uncertainty estimates from the repulsive ensembles are again very
small. The morhping-aware approach again performs worse.

Next, we evaluate the average log-likelihood ratio for the BSM event
sample in Fig.~\ref{fig:WZ_parton_2D_BSM}. The agreement with the true
likelihood ratio for the derivative learning is very good. Even though the parameter point is well
outside the expected $3\,\sigma$ exclusion region of LHC Run-3 (see
below) and features a sizeable enhanced tail in the $m_{WZ}^T$
distribution, the derivative learning yields stable results. The morphing-aware approach provides results close to the true likelihood ratio but sizeable deviations are visible.

We conclude that the considered scenario is local enough in the observable space for derivative learning to perform very well (see Sec.~\ref{sec:toy_compare}). The better phase-space coverage of the morphing-aware approach does not provide any benefit for the considered scenario.

%%%%%%%%%%%%%%%%%%%%%%%%%%%%%%%%%%%%%%%%%%%%%%%%%%%%%%%%%%%%%%%%%%%
\subsection{Reconstruction level}

For the reco-level analysis, we consider the leptonic $W$
and $Z$-decays and require exactly three leptons and no jet with
\begin{align}
  p_{T,\ell} > 15~\gev \qqquad 
  |\eta_\ell| < 2.5 \qqquad 
  p_{T,j} > 20~\gev \qqquad 
  |\eta_j| < 2.5 \; .
\end{align}
Before the likelihood inference, we impose the pre-selection cuts
\begin{align}
  m_Z^{\ell\ell} = 81.2~...~101.2~\gev \qquad 
  m_T^{W} > 30~\gev \qquad 
  \ETmiss > 45~\gev \qquad 
  p_T^{\ell W} > 20~\gev \; .
\end{align}
They ensure that the backgrounds are negligible for the likelihood
inference.
As shown in App.~\ref{app:backgrounds}, incorporating backgrounds into
the likelihood inference is straightforward.

For the reconstruction level we focus on derivative learning. As seen
at parton level, it yields reliable results for our $WZ$ analysis. This method is well motivated for our current analysis, but we want to point out that it could be insufficient for a full NLO analysis. The reason is that the phase space and theory parameter dependence can cause the appearance of large weights in underpopulated regions. In such a scenario, morphing aware likelihood learning could be a more stable method. Additionally, fractional smearing would be a useful tool to further tackle this issue.

For our $WZ$ reco-level analysis, we employ L-GATr in addition to repulsive-ensemble MLP results. The main input features are the three lepton 4-momenta and
$\ETmiss$. In addition, we provide the sum of the lepton charges, the
number of jets, and several reconstructed high-level observables:
$m_{Z}^{\ell\ell}$, the reconstructed $Z$ boson transverse momentum
$p_T^Z$, $p_T^{W\ell}$, $m_T^W$, and the transverse mass of the di-boson 
system $m_T^{WZ}$.

To generate the training and validation datasets, we employ fractional
smearing with the threshold $t = 0.5$ for each of the targets
$R_{i,ij}(z_p)$. The training dataset contains around 650k events, the
validation dataset 220k events, and the test dataset (generated
without fractional smearing), 200k events.

%--------------------------------
\begin{figure}[t!]
    \includegraphics[width=1\textwidth,trim={4cm 0cm 4cm 0cm},clip]{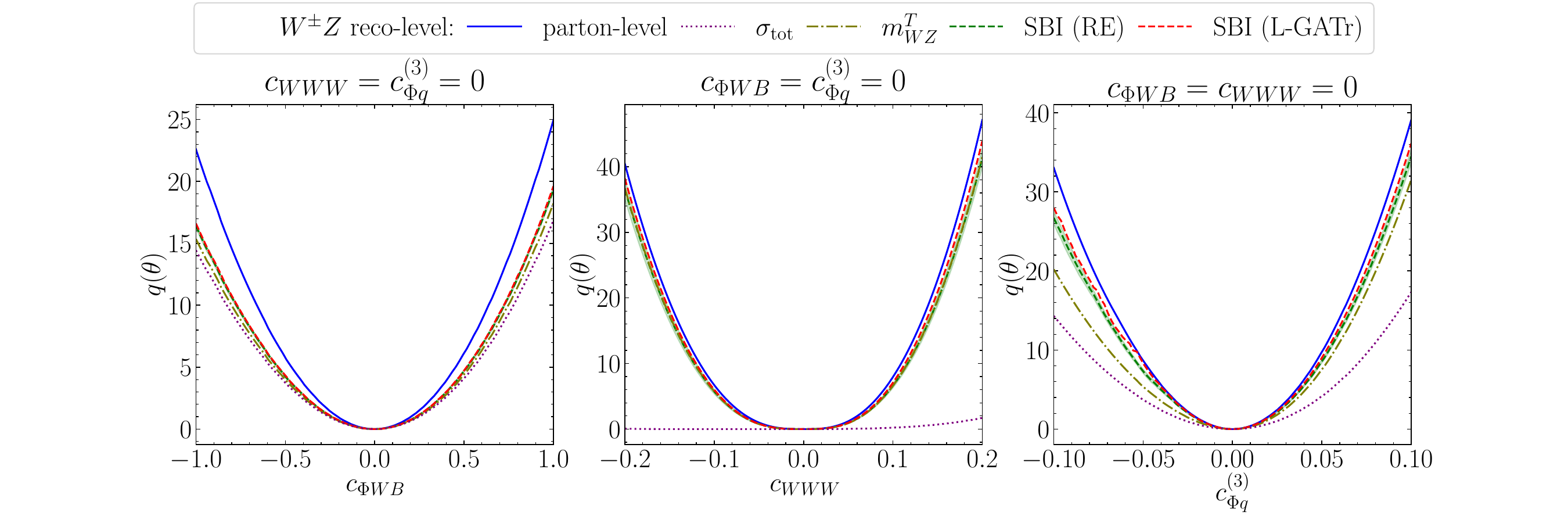}
    \caption{Test-statistics $q(\theta)$ for $\mathcal{L}= 300 \ifb$
      showing the parton-level truth, the reco-level SBI result using
      repulsive ensembles, and L-GATr. For comparison, we also show
      results based only on the cross-section and $m_{WZ}^T$. All
      methods are evaluated as functions of $c_{\Phi WB}$ (left),
      of $c_{WWW}$ (center), and of $c_{\Phi q}^{(3)}$ (right).}
    \label{fig:reco_WZ_1D}
\end{figure}
%--------------------------------

%--------------------------------
\begin{figure}[b!]
    \includegraphics[width=1\textwidth,trim={3.cm 0cm 4cm 0cm},clip]{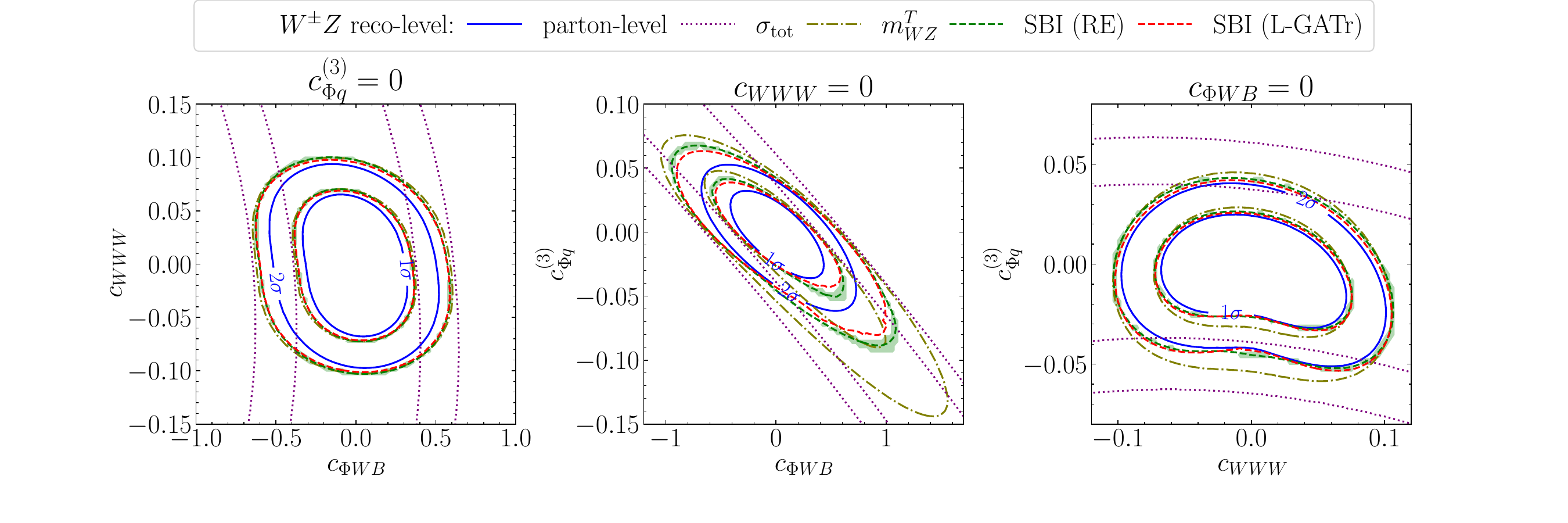}
    \caption{Negative log-likelihood for $\mathcal{L}= 300 \ifb$
      showing the parton-level truth,the reco-level SBI result using
      repulsive ensembles, and L-GATr. For comparison, we also show
      results based only on the cross-section and $m_{WZ}^T$. All
      methods are evaluated as a functions of $c_{\Phi WB}$ and
      $c_{WWW}$ (left), of $c_{\Phi WB}$ and $c_{\Phi q}^{(3)}$
      (center), and of $c_{WWW}$ and $c_{\Phi q}^{(3)}$ (right).}
    \label{fig:reco_WZ_2D}
\end{figure}
%--------------------------------

Fig.~\ref{fig:reco_WZ_1D} shows the results at reco-level for each
of the three Wilson coefficients.  In it, we replace the average
likelihood ratio with the test statistics $q(\theta)$ derived on the
basis of the full likelihood, see App.~\ref{app:like}.  For reference,
we also show the limits from the cross-section only and from the
1-dimensional $m_{WZ}^T$ histogram with seven bins and the boundaries
200~GeV, 400~GeV, 600~GeV, 800~GeV, 1~TeV, 1.5~TeV, and 2.5~TeV.  As
expected, the parton-level truth is the most constraining input. The
parton shower, the detector resolution, and the presence of missing energy result in a loss
information. At reco-level, limits from the total rate are poor, in
particular for $c_{WWW}$, which mainly affects the tail of the
$m_{WZ}^T$ distribution. Including kinematic information via the
$m_{WZ}^T$-histogram tightens the expected limits significantly. Using
SBI, the limits are further improved, most notably for $c_{\Phi
  q}^{(3)}$. We again find a very small uncertainty estimate from the
repulsive ensemble, indicating the robustness of our
results. L-GATr leads to a further improvement for $c_{WWW}$ and
$c_{\Phi q}^{(3)}$.

The 2-dimensional $1\,\sigma$ and $2\,\sigma$ confidence regions are shown in
Fig.~\ref{fig:reco_WZ_2D}.  The number of events is given by the
expected SM events at Run-3, with $\mathcal{L} = 300\ifb$. The cross-section-only limits have flat directions which are lifted by including kinematic information. As
expected, the histogram information is the least constraining,
followed by the repulsive ensemble SBI results. The most constraining
power at reco-level comes from L-GATr, with a substantial improvement
over the MLP along the diagonal in the $(c_{\Phi WB}, c_{\Phi
  q}^{(3)})$ direction.

%--------------------------------
\begin{figure}[t!]
    \centering
    \includegraphics[width=0.5\textwidth]{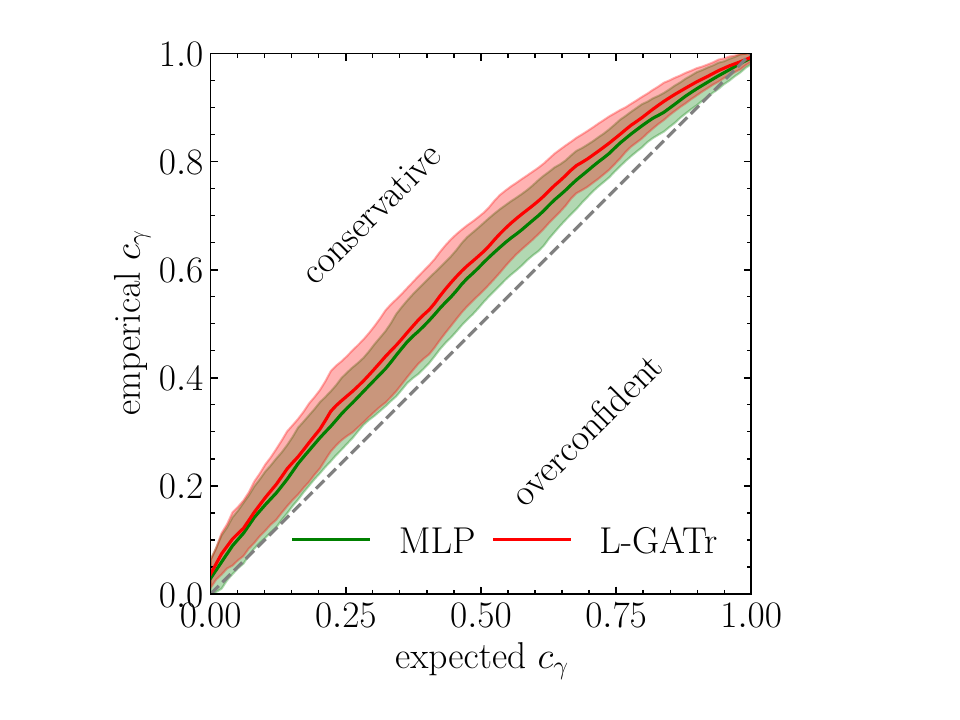}
    \caption{Expected coverage versus empirical coverage obtained
      scanning the likelihood in the full 3-dimensional
      $\theta$-space. The colored bands indicate the $1\,\sigma$
      uncertainty in the calculation of the empirical
      coverage.}
    \label{fig:reco_WZ_coverage}
\end{figure}
%--------------------------------

Finally, we show in Fig.~\ref{fig:reco_WZ_coverage} the empirical
coverage versus the expected coverage, described in
App.~\ref{app:like}. To evaluate the empirical coverage, we generate
random samples from the SM test dataset with the same event count as
the SM prediction. For each of these samples, we evaluate the
likelihood over the 3-dimensional $\theta$-space. We repeat this
procedure 500 times to determine the empirical coverage and its
statistical uncertainty.  We see that our learned likelihoods are
slightly conservative, meaning that the true SM point is within the
confidence region more often than expected. This shows that our
trained likelihood is not overestimating the sensitivity on the theory
parameters. Although it is not necessary in the present case, the
training loss of the neural networks can be adapted to ensure that the
likelihood estimation is not overconfident~\cite{Delaunoy:2022tbl}.

%%%%%%%%%%%%%%%%%%%%%%%%%%%%%%%%%%%%%%%%%%%%%%%%%%%%%%%%%%%%%%%%%%%
\section{Outlook}
\label{sec:outlook}

Simulation-based inference is the lead theme of modern LHC analyses, especially in view of the tenfold increase of data at the HL-LHC. It allows us to extract optimal information by comparing observed and simulated data without any low-dimensional or binned information bottleneck. To apply it systematically to LHC data, we need to employ modern machine learning, not only to improve the forward simulations, but also to get access to the unbinned likelihood (ratio).

In this paper we have targeted the likelihood ratio extraction and improved some of the standard methods and techniques~\cite{Brehmer:2019xox}. First, we have compared different ways to encode the perturbative physics structure into the likelihood learning and shown how uncertainties due to limited training data can be estimated using repulsive ensembles. A critical numerical improvement came from fractional smearing and weighted network training. To further improve the training towards complex scattering processes, we have employed a Lorentz-equivariant geometric algebra transformer, L-GATr. We have illustrated and studied all these technical improvements for a simple toy model, including a detailed comparison of the advantages and disadvantages of morphing-aware and derivative learning. 

For a physics application, we have looked at determining three Wilson coefficients simultaneously from $W^\pm Z$ production kinematics. At parton level, we confirmed that our approach is indeed optimal and extracts the true likelihood ratio over the entire relevant parameter space. The parton-level study served as a first benchmark for the novel morphing-aware approach. In the considered scenario, we found it to perform reasonably well but worse than the derivative learning method. At reco-level we produced numerically stable results, with a significant improvement over a rate measurement and a 1-dimensional histogram of the $WZ$-invariant mass. Our improved tool set will be the basis of a publicly available implementation as part of the next MadGraph/MadNIS release.

%%%%%%%%%%%%%%%%%%%%%%%%%%%%%%%%%%%%%%%%%%%%%%%%%%%%%%%%%%%%%%%%%%%
\section*{Acknowledgements}

We would like to thank Robert Sch\"ofbeck and Dennis Schwarz, the Vienna Gl\"uhwein workshop 2023, Johann Brehmer, and the L-GATr team, without whom this study would not have been successful. This research is supported through the KISS consortium
(05D2022) funded by the German Federal Ministry of Education and
Research BMBF in the ErUM-Data action plan, by the Deutsche
Forschungsgemeinschaft (DFG, German Research Foundation) under grant
396021762 -- TRR~257: \textsl{Particle Physics Phenomenology after the
  Higgs Discovery}, and through Germany's Excellence Strategy
EXC~2181/1 -- 390900948 (the \textsl{Heidelberg STRUCTURES Excellence
  Cluster}).  We would also like to thank the Baden-W\"urttem\-berg
Stiftung for financing through the program \textsl{Internationale
  Spitzenforschung}, pro\-ject \textsl{Uncertainties – Teaching AI its
  Limits} (BWST\_ISF2020-010). H.B.\ acknowledges support by the Alexander von Humboldt foundation. V.B.\ acknowledges financial support from the Grant No. ASFAE/2022/009 (Generalitat Valenciana and MCIN, NextGenerationEU PRTR-C17.I01).

\appendix
%%%%%%%%%%%%%%%%%%%%%%%%%%%%%%%%%%%%%%%%%%%%%%%%%%%%%%%%%%%%%%%%%%%
\section{Assessing the learned likelihood}
\label{app:like}

Since the actual likelihood is intractable (apart from toy models,
where we can calculate it analytically), the error of the learned
likelihood is also intractable. Nevertheless, it is possible to evaluate the
quality of the learned likelihood.

%%%%%%%%%%%%%%%%%%%%%%%%%%%%%%%%%%%%%%%%%%%%%%%%%%%%%%%%%%%%%%%%%%%
\subsection*{Uncertainties due to limited training data}

As a first aspect, we can assess the uncertainties due to limited
training data. This becomes especially important if we learn the
likelihood ratio only at one point using derivative learning and
then extrapolate from there. While this is a good approximation if the
distributions at a different parameter point cover the same phase
regions, this prescription necessarily breaks down for separated phase
distributions. A similar issue can occur for the morphing-aware setup
if interpolating between two benchmark points or if the likelihood is
evaluated at a parameter point far from all benchmark points.

To assess in which parameter region the learned likelihood ratio gives
precise results, we can learn the uncertainty due to limited training
data in a specific phase-space region as part of the network
training. In general, there are multiple ways to implement this: e.g.,
via Bayesian neural networks~\cite{Gal2016UncertaintyID,Bollweg:2019skg,Kasieczka:2020vlh,Bellagente:2021yyh,Butter:2021csz,Plehn:2022ftl} or repulsive
ensembles~\cite{DBLP:journals/corr/abs-2106-11642,Rover:2024pvr,Plehn:2022ftl}. Here, we use the repulsive ensemble approach, but
similar results are expected when using Bayesian neural networks.

A repulsive ensemble consists of a set of normal neural
networks which are trained in parallel. The loss function is adapted
to include a repulsive term which forces the networks to cover as much
of their weight space as possible while still converging on the actual
training objective. If the training sample is sparse in a certain
phase-space region, it becomes easier for the networks to provide a
reasonable fit. Due to the repulsive term, this results in a wider
spread between the networks. Eventually, the average of the networks
is used as the actual prediction of the ensemble, while their standard
deviation is used to estimate the uncertainty due to limited
training data.

If we use repulsive ensembles in the morphing-aware approach, we train one repulsive ensemble for every benchmark point. In the derivative approach, one repulsive ensemble is trained for each derivative w.r.t\ to the theory parameters. In both approaches, these ensembles are then used to build the total likelihood ratio (see Sec.~\ref{sec:likelihood_learning}). The estimated uncertainty of the total likelihood ratio can be obtained either
\begin{itemize}
    \item by error propagation of the uncertainties of the individual ensemble, or
    \item by computing the ensemble of the total likelihood ratio combining the ensembles of trained networks, and then computing the standard deviation over the resulting sample.
\end{itemize}
In the second case,  the selection of one network out of each ensemble can be randomized. Since the members of the ensemble are not ordered in any specific way, this choice does not make any difference in practice. If the number of ensemble members is large enough, both methods should give the same result. In the present work, we follow the second approach.

%%%%%%%%%%%%%%%%%%%%%%%%%%%%%%%%%%%%%%%%%%%%%%%%%%%%%%%%%%%%%%%%%%%
\subsection*{Limit setting and empirical coverage}
%\label{sec:limit_setting_and_coverage}

As an additional diagnostic tool, we can evaluate the statistical properties of the learned likelihood. The full likelihood for a set of events $\{x_i\}$ is given by
\begin{align}
    p_{\text{full}}(\{x\}|\theta) = \text{Pois}(n | \mathcal{L}\sigma(\theta)) \prod_i p(x_i | \theta) \;,
\end{align}
consisting out of the the total rate term $\text{Pois}(n | \mathcal{L}\sigma(\theta))$ with the luminosity $\mathcal{L}$ and the kinematic likelihoods for each event $p(x_i|\theta)$. Here, $\text{Pois}(k|\lambda) = \lambda^k e^{-\lambda}/k!$ is the Poisson distribution. We denote the corresponding full likelihood ratio as $r_\text{full}(\{x\}|\theta,\theta_0)$.

Using the full likelihood, we construct the test statistics
\begin{align}
    q(\theta) = -2 \log r_\text{full}(\{x\}|\theta, \hat\theta) = -2\left(\log r_\text{full}(\{x\}|\theta,\theta_0) - \log r_\text{full}(\{x\}|\hat\theta,\theta_0)\right)\;,
\end{align}
where $\hat\theta$ is its minimum which we estimated via
\begin{align}
    \hat\theta = \underset{\theta}{\text{argmax}}\log r_\text{full}(\{x\}|\theta,\theta_1).
\end{align}
In the asymptotic limit, the distribution $p(q(\theta)|\theta)$ is a chi-squared distribution. We can then calculate the $p$-value
\begin{align}
    p_\theta = \int_{q_\text{obs}(\theta)}^{\infty}dq\, p(q(\theta)|\theta) = 1 - F_{\chi^2}(q_\text{obs}(\theta)|k)\;,
\end{align}
which measures the confidence with which we can reject $\theta$. Here, $F_{\chi^2}(x|k)$ is the cumulative chi-squared distribution function with $k$ degrees of freedom. $q_\text{obs}(\theta)$ is the observed value for $q(\theta)$. Using this, the $\gamma$ confidence region is defined by all $\theta$ values for which $p_\theta < \gamma$.

As a consistency check for our learned likelihood ratio $r_\varphi(x|\theta,\theta_0)$, we can generate $n$ samples for a given $\theta$ and check for how many of the $n$ samples the true $\theta$ is contained in a given $\gamma$ confidence region. This defines the coverage
\begin{align}
    c_\gamma \equiv \left\langle \mathbb{1}\left(p_{\theta_0}(\{x\}) > 1-\gamma\right) \right\rangle_{\{x\}}\;,
\end{align}
where $\mathbb{1}$ is the indicator function, which evaluates to one if the condition in the brackets is true and to zero otherwise. The average value is taken over the $n$ samples. The coverage is exactly $\gamma$ if the likelihood estimation is perfect. This means the true minimum is within the $\gamma$ confidence level in a fraction $\gamma$ of all cases. If the fraction is higher --- meaning that the empirical $c_\gamma > \gamma$ ---, our learned likelihood is conservative or underconfident. If it is lower --- meaning that the empirical $c_\gamma < \gamma$ ---, the learned likelihood is overconfident.

%%%%%%%%%%%%%%%%%%%%%%%%%%%%%%%%%%%%%%%%%%%%%%%%%%%%%%%%%%%%%%%%%%%
\section{Fractional-smearing datasets}
\label{sec:frac_smearing_datasets}

%--------------------------------
\begin{figure}
    \centering
   \includegraphics[width=0.49\textwidth,trim={.5cm 0cm .5cm 0cm},clip]{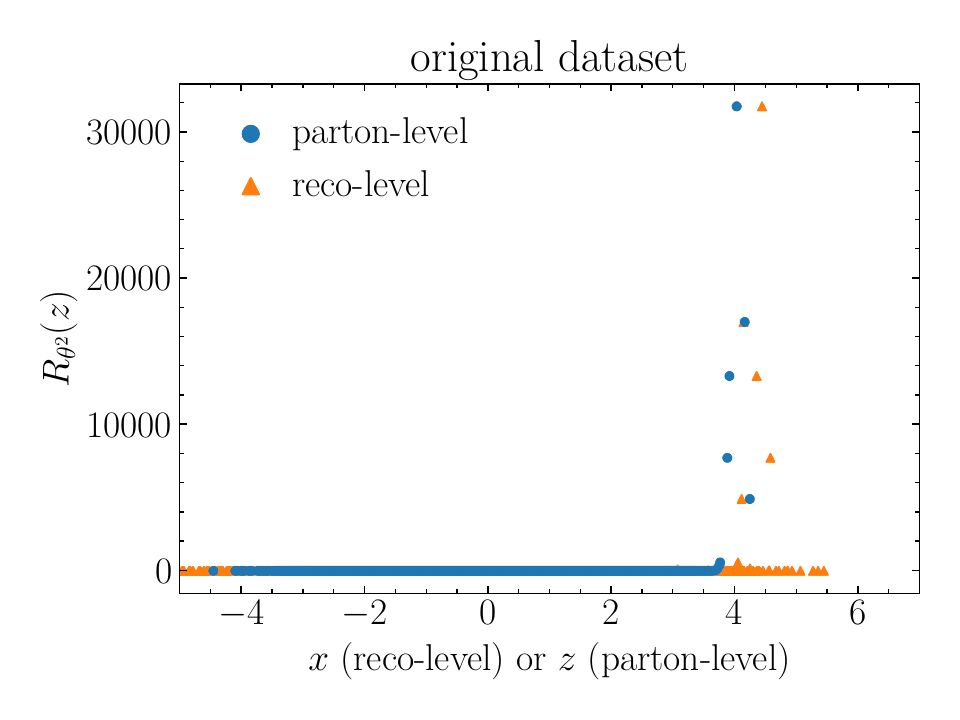}
  \includegraphics[width=0.49\textwidth,trim={.5cm 0cm .5cm 0cm},clip]{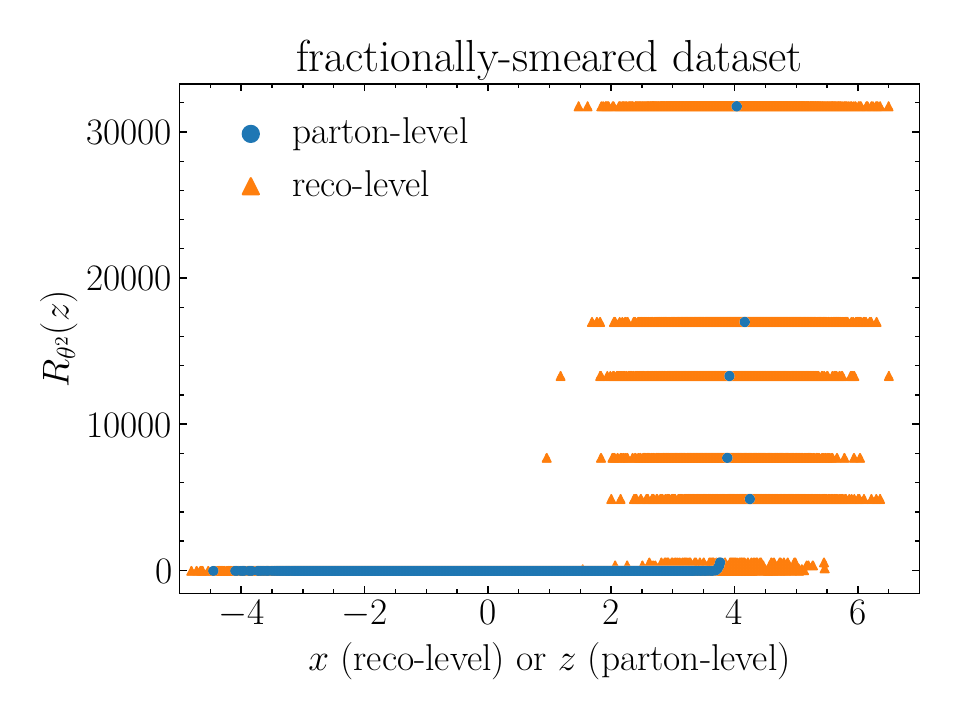}
    \caption{Left: Second derivative of the differential cross-section with respect to $\theta$ as a function of $x$ for the reco-level points or $z$ for the parton-level points. Shown is the parton-level (blue) and the reco-level (orange) distribution for the original dataset. Right: Same as left but for the fractionally-smeared dataset.}
    \label{fig:fractional_smearing_dataset_comparison}
\end{figure}
%--------------------------------

This original dataset before fractional smearing is shown in the left panel of Fig.~\ref{fig:fractional_smearing_dataset_comparison}. This Figure shows the second derivatives of either the parton-level differential cross-section or the reco-level differential cross-section w.r.t.\ $\theta$ as a function of either $x$ for the reco-level points or $z$ for the parton-level points. The blue points show the parton-level distribution of events; the orange points, the reco-level distribution. 

Looking at Fig.~\ref{fig:fractional_smearing_dataset_comparison}, the issue with the network training becomes immediately obvious. The parton-level events (blue circles) are smeared randomly to the left or right to obtain the reco-level training dataset (orange triangles). Since the event density in the high $x$ or $z$ region is low, this smearing can result in a random bias, making it very hard for the network to converge towards the true distribution.

This issue is alleviated by the fractional smearing procedure, as illustrated in the right panel of Fig.~\ref{fig:fractional_smearing_dataset_comparison}. Now every event with a high target is repeatedly smeared resulting in a distribution of smeared events (orange triangles) around each parton-level event (blue circles). This makes it easier for the network to learn the reco-level distribution. 

%%%%%%%%%%%%%%%%%%%%%%%%%%%%%%%%%%%%%%%%%%%%%%%%%%%%%%%%%%%%%%%%%%%
\section{Morphing-aware likelihood estimation for \texorpdfstring{$pp\to W^\pm Z$}{pp->WZ}}
\label{app:morphing-aware LHC}

For the morphing-aware estimation of the $pp\to W^\pm Z$ likelihood at parton level, we choose the following basis points
\begin{align}
    \label{eq:morphing_basis_points}
    \theta_1&=(-4,0,0),      \qquad &\theta_2 &=(4,0,0);\notag\\
    \theta_3&=(0,-0.2,0),    \qquad &\theta_4 &=(0,0.2,0);\notag\\
    \theta_5&=(0,0,-0.2),    \qquad &\theta_6 &=(0,0,0.2);\notag\\
    \theta_7&=(-1.2,-0.09,0), \qquad &\theta_8 &=(-1.2,0,-0.09);\notag\\
    \theta_9&=(0,-0.09,-0.09)\,, &&
\end{align}
where $\theta=(c_{\Phi WB},c_{WWW},c_{\Phi q}^{(3)})$. We find that choosing basis points along the coordinate axis yields more stable results due to a simplification of the morphing matrix inversion.

Moreover, we choose a different training loss to the one used in Sec.~\ref{sec:toy}. Instead of the BCE loss defined in Eq.\ref{eq:BCE_ALICE_loss}, we use an MSE-based loss, which we find to yield more stable results. In particular, the loss reads
\begin{align}
    \loss = 
    \XXLangle \left[ r(z_{p,0}|\theta_i,\theta_0)\right.&-\left.r_\varphi(x_0|\theta_i,\theta_0)\right]^2+
    \notag \\
    & + \left[ \frac{1}{r(z_{p,i}|\theta_i,\theta_0)}- \frac{1}{r_\varphi(x_i|\theta_i,\theta_0)} \right]^2\XXRangle_{x_i,z_{p,i} \sim p(x|z_p)p(z_p|\theta_i)} \; ,
    \label{eq:MSE_loss_morphing}
\end{align}
where the first term is only evaluated for events from the denominator hypothesis while the second term is only evaluated for events from the numerator hypothesis. The inversion of the ratio in the second term ensures convergence to the correct true ratio limit.

%%%%%%%%%%%%%%%%%%%%%%%%%%%%%%%%%%%%%%%%%%%%%%%%%%%%%%%%%%%%%%%%%%%
\section{Backgrounds}
\label{app:backgrounds}

In the presence of backgrounds, the squared matrix element can be written in the form 
\begin{align}
    \mat^2(z_p|\theta) = \mat^2_\sig(z_p|\theta) + \underbrace{2\text{Re}\left[\mat_\sig(z_p|\theta) \mat_\bkg^*(z_p)\right]}_{\text{intf}} + \mat^2_\bkg(z_p)\;,
\end{align}
where $\mat_\sig(z_p|\theta)$ is the matrix element of the signal process and $\mat_\bkg(z_p)$ is the matrix element of the background process, which does not depend on $\theta$.

If the signal and background do not have the same partonic final state, the interference term is zero. In this case, we can split up the differential reco-level cross-section,
\begin{align}
    d\sigma(x|\theta) = d\sigma_\sig(x|\theta) + d\sigma_\bkg(x)\;,
\end{align}
and also the likelihood,
\begin{align}
    p(x|\theta) =  \frac{\sigma_\text{sig}(\theta)}{\sigma_\text{sig}(\theta) + \sigma_\text{bkg}}p_\text{sig}(x|\theta) + \frac{\sigma_\text{bkg}}{\sigma_\text{sig}(\theta) + \sigma_\text{bkg}} p_\bkg(x)\;.
\end{align}
Then, we can write the likelihood ratio in the form
\begin{align}
    r(x|\theta,\theta_0) 
    &= \frac{\sigma_\sig(\theta_0) + \sigma_\bkg}{\sigma_\sig(\theta) + \sigma_\bkg}\frac{1 + \frac{\sigma_\sig(\theta)}{\sigma_\bkg}\frac{p_\sig(x|\theta_0)}{p_\bkg(x)}\frac{p_\sig(x|\theta)}{p_\sig(x|\theta_0)}}{1 + \frac{\sigma_\sig(\theta_0)}{\sigma_\bkg}\frac{p_\sig(x|\theta_0)}{p_\bkg(x)}} 
\end{align}
We can now separately learn a background--signal classifier at the point $\theta_0$ to extract the ratio $p_\text{sig}(x|\theta_0)/p_\text{bkg}(x)$. Then, the signal-only likelihood ratio $p_\text{sig}(x|\theta)/p_\text{sig}(x|\theta_0)$ can be learned separately using the methods outlined in the main body of the text. A similar approach has already been applied in Ref.~\cite{Mastandrea:2024irf}.

%%%%%%%%%%%%%%%%%%%%%%%%%%%%%%%%%%%%%%%%%%%%%%%%%%%%%%%%%%%%%%%%%%%
\bibliography{tilman,refs}
\end{document}